\documentclass[lettersize,journal]{IEEEtran}
\usepackage{amsmath,amsfonts}
\usepackage{algorithmic}
\usepackage{algorithm}
\usepackage{array}
\usepackage[caption=false,font=normalsize,labelfont=sf,textfont=sf]{subfig}
\usepackage{textcomp}
\usepackage{stfloats}
\usepackage{url}
\usepackage{verbatim}
\usepackage{graphicx}
\usepackage{cite}
\usepackage{amssymb}
\usepackage{xcolor}
\usepackage{epsfig}
\usepackage{epstopdf} 
\usepackage{multirow}
\usepackage{arydshln}
\usepackage{booktabs}
\usepackage{makecell}
\usepackage{empheq}
\begin{document}

\title{Dynamic Task and Resource Scheduling Towards Space-Air-Ground-Sea Integrated Network}

 




\author{Yufei Ye,~\IEEEmembership{Graduate Student Member,~IEEE}, Shijian Gao,~\IEEEmembership{Member,~IEEE}, Xinhu Zheng,~\IEEEmembership{Member,~IEEE}, and Liuqing Yang,~\IEEEmembership{Fellow,~IEEE}
\thanks{Yufei Ye is with the Intelligent Transportation Thrust, The Hong Kong University of Science and Technology (Guangzhou), Guangzhou 511458, China (e-mail: yye760@connect.hkust-gz.edu.cn).}
\thanks{Shijian Gao is with the Internet of Things Thrust, The Hong Kong University of Science and Technology (Guangzhou), Guangzhou 511458, China (e-mail: shijiangao@hkust-gz.edu.cn).}
\thanks{Xinhu Zheng and Liuqing Yang are with the Intelligent Transportation Thrust and Internet of Things Thrust, The Hong Kong University of Science and Technology (Guangzhou), Guangzhou 511458, China (e-mail: xinhuzheng@hkust-gz.edu.cn; lqyang@hkust-gz.edu.cn).}
}



\maketitle

\begin{abstract}
In the context of 6G ubiquitous connectivity, the space–air–ground–sea integrated network (SAGSIN) emerges as a new paradigm for pervasive service provisioning. To support expanding maritime activities in infrastructure-scarce ocean areas, we propose an innovative dynamic task and resource scheduling approach for SAGSIN to deliver computing services for vessels. It integrates broad-coverage satellites, relay-capable high-altitude platform (HAP), energy-sufficient coastal base station (BS), and flexibly deployed uncrewed aerial vehicles (UAVs) to accommodate wide-area, highly mobile, and sustained maritime services. To address the challenge of task scheduling across four layers, a dynamic task offloading algorithm is developed. It steers task flows toward servers with light loads, strong computing capabilities, and high-rate links based on real-time system states to reduce task execution delay, integrating an anticipatory satellite handover strategy to mitigate post-handover congestion and improving satellite resource utilization. Considering the limited endurance of UAVs, we impose residual energy constraints to ensure task backlog handover and safe return. Furthermore, the UAV-BS bandwidth allocation, UAV trajectories, and computing resource allocation are jointly optimized to enhance the connectivity among low‑altitude devices and accelerate task completion. Simulation results validate the proposed method’s superior adaptability to system resource variations during task execution in complex maritime environments, achieving at least a 23\% reduction in average task delay over benchmarks.

\end{abstract}


\section{Introduction}
\IEEEPARstart{T}{he} Sixth Generation (6G) networks are envisioned to enable ubiquitous computing and connectivity across heterogeneous environments \cite{majamshed2025non}. Under this vision, rapidly expanding maritime activities, such as ocean resource exploration and environmental monitoring, are imposing increasing demands on communication and computing resources. However, the limited onboard computing power of vessels and scarce infrastructures in ocean areas pose critical challenges for supporting emerging maritime applications \cite{jyou2025joint}.

Maritime service provisioning typically requires wide-area coverage and continuous service availability. Therefore, satellites with broad coverage \cite{zhuang2026joint, zwang2025two} and coastal base stations (BSs) \cite{zhuang2026two} with stable energy supply have become the major service providers in ocean areas. Nevertheless, the long communication distances and severe path loss of satellite links lead to significant delay, while the limited coverage of BSs restricts their capability to support mobile vessels and distant offshore areas. With the agile mobility and rapid deployment, uncrewed aerial vehicles (UAVs), as key enablers of the low‑altitude economy (LAE), have attracted considerable attention as a means of providing computing services to mobile onboard devices \cite{yye2026multitier, ybai2025dynamic}. To extend the capabilities of UAVs to ocean areas, recent studies have explored the spectrum-efficient maritime task offloading \cite{yhe2025delay}, the discrete-continuous hybrid framework for task offloading decision and communication resource allocation \cite{jning2025marho}, the sea buoy-UAV cooperative computing architecture \cite{wxu2025madrl}, and the energy-minimization offloading scheme for vessels and underwater sensors \cite{mdai2023uav}. Despite these advances, the compact physical sizes of UAVs result in limited computing capabilities and endurance, which hinders their ability to handle intensive computational demands \cite{sqi2024minimizing}.

In this context, the space‑air‑ground‑sea integrated network (SAGSIN) emerges as a new network architecture to achieve ubiquitous connectivity \cite{sgao2026integrated}. For maritime service support, SAGSIN exploits the complementary advantages of heterogeneous servers, including broad-coverage satellites, energy-sufficient BSs, and flexibly deployed UAVs, and integrates their resources to provide computing services for maritime devices \cite{xwang2025bridging}. To harness the potential of this integrated architecture, recent efforts have been devoted to collaborative edge computing and resource orchestration in SAGSIN. When UAVs are assisted by the BS, maritime task delay is minimized through the joint optimization of communication-computing resources with task offloading ratios \cite{mdai2023latency} and service caching decisions \cite{yzhang2025joint}. The authors in \cite{mli2026online} develop an online task offloading framework to cope with the time-varying channel conditions. Leveraging the powerful computing capabilities of low earth orbit (LEO) satellites, UAV‑satellite collaborative computing frameworks have been designed to save energy for the entire system \cite{zwang2025double} or UAV servers \cite{sjung2023marine}. A scheme to support tasks with different delay sensitivities is proposed in \cite{sqi2025joint}, and a satellite‑assisted harvesting‑and‑offloading method is developed to prolong UAV endurance in \cite{mdai2025energy}. Moreover, \cite{hzhang2025energy} designs a three‑tier computing system that augments satellites with high access demands by BS, while a multi-armed bandit-based algorithm is proposed in \cite{tyang2022multi} to jointly reduce task delay and energy cost. Acting as relays bridging the space layer and low-altitude region, the high‑altitude platforms (HAPs) can effectively alleviate the high path loss of satellite links while compensating for resource-limited UAVs. Multi-HAP-assisted computing architectures for SAGSIN have been recently investigated to further reduce task completion time \cite{wwu2025multi} and hybrid service cost \cite{wli2026efficient}. Meanwhile, \cite{zlin2024maritime} improves the energy efficiency of heterogeneous servers and \cite{dwang2023double} investigates the secure task offloading in SAGSIN.



Despite the great potential of SAGSIN, several critical challenges remain insufficiently addressed. To cope with the high complexity of task offloading across four layers, most existing studies reduce the feasible decision space by predefining device associations between certain layers, which limits the flexibility of task and resource scheduling and may lead to load imbalance. Meanwhile, the static one-shot scheduling strategies fail to track evolving system conditions during task transmission, leading to degraded performance when tasks reach destination servers. Moreover, adaptive strategies for handling abrupt satellite state transition caused by handover remain underexplored. Satellite offloading decisions based solely on the state of current satellite may result in post-handover congestion or resource underutilization. In addition, existing strategies for UAV energy management mainly focus on minimizing energy consumption or imposing fixed energy budgets, which may compromise service quality. Ensuring that the real-time residual energy of UAVs can support task backlog handover and safe return is more robust for efficient and sustainable service provisioning in maritime environments.

To holistically address these challenges, in this paper, we propose a novel \underline{d}ynamic t\underline{a}sk and re\underline{s}ource sc\underline{h}eduling method (DASH) for SAGSIN to provide efficient computing support for vessels. DASH enables flexible adaptation to system dynamics by introducing a low‑complexity, layer‑wise dynamic task offloading algorithm inspired by backpressure routing theory. It adaptively adjusts fine‑grained task scheduling decisions according to multi-dimensional, real‑time system conditions, including available computing resources of heterogeneous servers, task congestion levels, link capacities, and network topology. To address the underexplored handover problem, DASH further incorporates an anticipatory strategy that proactively regulates satellite offloading traffic based on the joint states of both the current and incoming satellites, thereby mitigating post‑handover congestion while improving satellite resource utilization. Beyond task offloading, we jointly optimize UAV‑BS bandwidth allocation, UAV trajectory planning, and computing resource allocation. This strengthens the connectivity between vessels and low‑altitude UAVs and boosts task completion through the co‑design of resource scheduling and physical network architecture. Considering the limited UAV endurance and impracticality of emergency landing in maritime environment, we incorporate UAV residual energy constraints to maintain sufficient energy for task backlog handover and safe return to charging station. This design ensures UAV operational safety and service continuity while avoiding unnecessary energy restrictions that may compromise service quality. Through adaptive multi‑layer load balancing and demand‑driven resource allocation, DASH provides efficient computing services while improving resource utilization efficiency and the ecological stability of the system, laying a crucial foundation for sustainable SAGSIN development. Simulation results demonstrate that DASH achieves superior adaptability to variations in computing and bandwidth resources, substantially reducing the average task execution delay compared with benchmarks while effectively mitigating post-handover satellite congestion. The main contributions of this work are summarized as follows:

\begin{enumerate}
    \item We propose a novel dynamic task and resource scheduling approach for SAGSIN. A low-complexity layer-wise task offloading scheme that adapts to multi-dimensional system dynamics is developed, integrating an anticipatory satellite handover strategy to mitigate post-handover congestion and improve resource utilization.
    
    \item Beyond task offloading, we jointly optimize UAV-BS bandwidth allocation, UAV trajectories, and computing resource allocation to further accelerate task completion. Moreover, UAV residual energy constraints are incorporated to ensure task backlog handover and safe return, thereby maintaining UAV operational safety and service continuity in maritime environments.
    
    \item Extensive experiments validate the superiority of the proposed method over benchmarks in adapting to varying system resources, achieving at least a 23\% reduction in average task delay while effectively alleviating post-handover congestion on satellite and balancing workloads across heterogeneous servers.
\end{enumerate}

The rest of this article is organized as follows. In Section II, we present the system model of the proposed SAGSIN architecture. Then we formulate the overall problem in Section III. In Section IV, we detail the anticipatory satellite handover strategy and multi-layer task offloading scheme. The solution for joint optimization of multifaceted resources and the complexity analysis of overall proposed approach are elaborated in Section V. We illustrate and discuss the simulation results in Section VI and finally conclude this paper in Section VII.


\section{System Model}

In this section, we present the system model of the proposed SAGSIN architecture, which consists of the network model, communication model, computation and task queue models, and UAV model, providing a unified framework for subsequent problem formulation and algorithm design.

\subsection{System Overview}
\begin{figure}[!t]
\centering
\includegraphics[width=3.4 in]{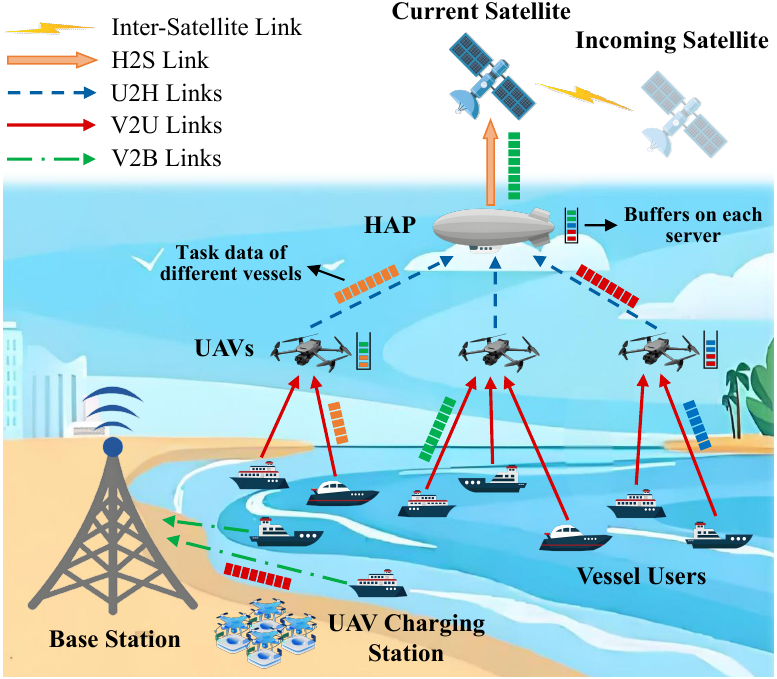}
\caption{Illustration of the multi-layer network architecture for SAGSIN.}
\label{network_model}
\end{figure}

Fig.~\ref{network_model} depicts the proposed SAGSIN system architecture. Specifically, the sea layer consists of $V$ vessel users denoted by set $\mathcal{V}=\{1,2, ..., v, ...,V\}$, while a coastal base station (BS) server is deployed in the ground layer. The air layer comprises $U$ UAVs operating in low-altitude regions as lightweight aerial servers, represented by set $\mathcal{U}=\{1,2, ..., u, ...,U\}$, and an HAP functioning as an aerial BS server in high-altitude region. The space layer consists of an LEO satellite server currently serving the considered area and an incoming satellite server that will subsequently provide service to this area. Given the limited endurance of UAVs and the impracticality of emergency landings caused by low battery levels in maritime environments, we deploy a UAV charging station. To characterize the system dynamics, we discretize time into slots, each indexed by $n$ starting from 0 and having a length of $\tau$. We take sea surface as the reference plane. In time slot $n$, the horizontal locations of vessel user $v$, UAV $u$, coastal BS, HAP, and charging station are denoted by $\boldsymbol{W_v}(n) = (x_v(n), y_v(n))$, $\boldsymbol{W_u}(n) = (x_u(n), y_u(n))$, $\boldsymbol{W_b} = (x_b, y_b)$, $\boldsymbol{W_h}(n) = (x_h(n), y_h(n))$, and $\boldsymbol{W_{cs}} = (x_{cs}, y_{cs})$, respectively. The height of UAVs, coastal BS, and HAP are signified by $H_u$, $H_b$, and $H_h$, respectively.

We denote the size of task data generated by vessel $v$ as $D_v$ bits and its computational density as $C_v$ cycles/bit, which indicates the number of CPU cycles required to process each bit of data. The operation process of the whole system is described as follows: In each time slot, each vessel user selects either a UAV or the coastal BS based on the real-time multi-dimensional system states to offload a portion of its task data. Each UAV and the HAP offload part of the selected vessel’s task data stored in their local buffers to the HAP and satellite server, respectively. Meanwhile, the coastal BS, UAVs, HAP, and satellite server compute portions of buffered task data from different vessels. For clarity and conciseness, the key notations in this work and their definitions are summarized in Table~\ref{tab:notations}, where the index of the time slot is omitted.

\begin{table}[!t]
\caption{Main Notations and Definitions}
\label{tab:notations}
\centering
\setlength{\arrayrulewidth}{1pt} 
\begin{tabular}{!{\vrule width \arrayrulewidth} m{1.5cm}<{\centering} !{\vrule width \arrayrulewidth} m{6.2cm}<{\centering} !{\vrule width \arrayrulewidth}}
\Xhline{1pt}
\textbf{Notation} & \textbf{Definition}\\
\Xhline{1pt}
$\mathcal{V},\mathcal{U}$ & Set of vessel users and UAVs, respectively\\
\Xhline{0.5pt}
$D_v, C_v$ & Size and computational density of task data generated by vessel $v$, respectively \\
\Xhline{0.5pt}
$\tau$ & Duration of each time slot\\
\Xhline{0.5pt}
$T_v$ & Total execution delay of vessel $v$'s task\\
\Xhline{1pt}
$\boldsymbol{W_v},\boldsymbol{W_u},$ $\boldsymbol{W_b}, \boldsymbol{W_h}$ & Positions of vessel $v$, UAV $u$, coastal BS, and HAP, respectively\\
\Xhline{0.5pt}
$H_u, H_b, H_h$ & Height of UAVs, coastal BS, and HAP, respectively\\
\Xhline{0.5pt}
$d_{h,s}$ & Distance between HAP and satellite \\
\Xhline{0.5pt}
$d_{safe}$ & Minimum safety distance among UAVs\\
\Xhline{0.5pt}
$S_{u,max}$ & Maximum flight speed of UAV $u$\\
\Xhline{1pt}
$\boldsymbol{a_{v,u}}$ & V2U offloading decision indicators of vessel $v$\\
\Xhline{0.5pt}
$a_{v,b}$ & V2B offloading decision indicator of vessel $v$\\
\Xhline{0.5pt}
$\boldsymbol{a_{v,u,h}}$ & U2H offloading decision indicators of UAV $u$\\
\Xhline{0.5pt}
$\boldsymbol{a_{v,h,s}}$ & H2S offloading decision indicators of HAP\\
\Xhline{1pt}
$B_{v,b}, B_{v,u}$ & Bandwidth allocated by BS and UAV $u$ to vessel $v$\\
\Xhline{0.5pt}
$B_{h}, B_{s}$ & Bandwidth for U2H and H2S communications\\
\Xhline{0.5pt}
$R_{v,u}$ & Data rate between vessel $v$ and UAV $u$ \\
\Xhline{0.5pt}
$R_{v,b}$ & Data rate between vessel $v$ and coastal BS \\
\Xhline{0.5pt}
$R_{u,h}$ & Data rate between UAV $u$ and HAP \\
\Xhline{0.5pt}
$R_{h,s}$ & Data rate between HAP and satellite\\
\Xhline{1pt}
$F_{v,i}$ & Computing resource allocated by server $i$ to the task data of vessel $v$ \\
\Xhline{0.5pt}
$F_{i}$ & Available computing resource of server $i$\\
\Xhline{0.5pt}
$D_{v,i}^{comp}$ & Amount of vessel $v$'s data computed by server $i$\\
\Xhline{1pt}
$Q_{v,0}$ & Amount of task data on local device of vessel $v$ \\
\Xhline{0.5pt}
$Q_{v,i}$ & Amount of task data of vessel $v$ on server $i$ \\
\Xhline{1pt}
$E_u^{cons}$ & Total energy consumption of UAV $u$ in each time slot\\
\Xhline{0.5pt}
$E_u$ & Remaining energy of UAV $u$ at the beginning of each time slot \\
\Xhline{0.5pt}
$E_u^{cs}$ & Energy required for UAV $u$ to return to charging station at the beginning of each time slot \\
\Xhline{0.5pt}
$E_u^{re}$ & Energy required for UAV $u$ to transmit remaining data to HAP at the beginning of each time slot \\
\Xhline{1pt}
\end{tabular}
\end{table}


\subsection{Communication Model}

\subsubsection{Vessel-to-UAV and Coastal BS Communication}

In light of the characteristics of marine environment, e.g., strong Line-of-Sight (LoS) links, the wireless channel in ocean area is primarily affected by wave-induced multipath effects and weather conditions. According to \cite{mdai2023latency, zwang2025double, sqi2025joint}, the vessel-to-UAV (V2U) link and vessel-to-BS (V2B) link are modeled as the combination of large-scale path loss and small-scale Rician fading. The path loss and the Rician fading between vessel $v$ and UAV $u$ or BS are respectively given by
\begin{equation*}
h^L_{v,i}(n)=\frac{L_0}{d_{v,i}(n)^2}=\frac{L_0}{{H_i}^2+\| \boldsymbol{W_i}(n) - \boldsymbol{W_v}(n)\|^2}, \tag{1a}
\end{equation*}
\begin{equation*}
h^R_{v,i}(n)=\sqrt{\frac{K_0}{1+K_0}}+\sqrt{\frac{1}{1+K_0}}o_{v,i}(n), \tag{1b}
\end{equation*}
where $i \in \{u,b\}$, $d_{v,i}(n)$ denote the distance between vessel $v$ and UAV $u$ or BS at time slot $n$, $L_0$ is the reference path loss at distance $d_0 = 1$m, $o_{v,i}(n) \in \mathcal{CN}(0,1)$, and $K_0$ is Rician factor. Employing orthogonal frequency division multiple access (OFDMA) technology \cite{mdai2023latency, wwu2025multi}, the transmission data rate between vessel $v$ and UAV $u$ or BS is expressed as
\begin{equation*}
R_{v,i}(n) = B_{v,i}(n) \log_2 \left(1+\frac{P_v(n)h_{v,i}(n)}{N_0 B_{v,i}(n)}\right), \tag{2}
\end{equation*}
where $h_{v,i}(n)=h_{v,i}^{L}(n)|h_{v,i}^{R}(n)|^{2}$ is the channel gain, $B_{v,i}(n)$ signifies the bandwidth allocated to each V2U or V2B channel, $P_v(n)$ denotes the transmit power of vessel $v$, and $N_0$ is noise power spectral density.

\subsubsection{UAV-to-HAP Communication}

The total path loss between UAV $u$ and HAP at time slot $n$ is formulated as \cite{wwu2025multi}
\begin{equation*}
L_{u,h}(n)=L_{f,u,h}(n)+L_g+L_s, \tag{3}
\end{equation*}
where $L_g$ denotes the atmospheric attenuation and $L_s$ represents the attenuation caused by ionospheric or tropospheric scintillation. $L_{f,u,h}(n)$ is the free space path loss, which is given by $L_{f,u,h}(n)=32.45+20\log_{10}f_c+20\log_{10}d_{u,h}(n)$, where $f_c$ is the carrier frequency in MHz and $d_{u,h}(n)=\sqrt{(H_h - H_u)^2+\| \boldsymbol{W_h}(n) - \boldsymbol{W_u}(n)\|^2}$ denotes the distance between UAV $u$ and HAP in km. Hence the transmission rate of this UAV-to-HAP (U2H) link is expressed as
\begin{equation*}
R_{u,h}(n)=B_h\log_2\left(1+\frac{P_u(n) G_u G_h 10^{-\frac{L_{u,h}(n)}{10}}}{ N_0 B_h}\right), \tag{4}
\end{equation*}
where $P_u(n)$ is UAV transmit power, $G_u$ and $G_h$ signify the antenna gain of UAV $u$ and HAP, respectively, and $B_h$ is the bandwidth for U2H communication. Denoting the data volume of the selected vessel $v^*$ that UAV $u$ offloads to HAP in time slot $n$ as $D_{v^*,u,h}^{tx}(n)$, the corresponding transmission energy of this UAV is $E_u^{tx} (n) = P_u (n)D_{v^*,u,h}^{tx}(n)/R_{u,h}(n)$.

\subsubsection{HAP-to-Satellite Communication}

Following \cite{wwu2025multi}, the large-scale fading between HAP and LEO satellite can be captured by
\begin{equation*}
\eta_{h,s}(n)=\hat{\mu}+10\mu\log_{10}\left(\frac{d_{h,s}(n)}{d_0}\right)+\delta, \tag{5}
\end{equation*}
where $\hat{\mu}$ denotes the intercept parameter (i.e., path loss at reference distance), $\mu$ is the path loss exponent, $d_{h,s}(n)$ is the HAP-to-satellite (H2S) distance at time slot $n$, and $\delta$ is the model deviation in fitting represented by a zero-mean Gaussian random variable with standard deviation $\omega = 0.1$ \cite{sshassan2021blue}. The transmission rate of the H2S link is given by
\begin{equation*}
R_{h,s}(n)=B_s \log_2\left(1+\frac{P_h(n) h_{h,s}(n)}{N_0 B_s}\right), \tag{6}
\end{equation*}
where $B_s$ denotes the bandwidth for H2S communication and $P_h(n)$ is the transmit power of HAP. The channel gain is formulated as $h_{h,s}(n)=K_0 10^{-\frac{\eta_{h,s}(n)}{10}} G_h G_s$, where $G_s$ is the antenna gain of satellite.

\subsubsection{Inter-Satellite Communication}

When the current satellite is about to leave the considered region with unfinished computation tasks, it forwards the remaining backlogged task data to the incoming satellite via the inter-satellite link (ISL) to enable seamless service handover. Benefiting from the abundant optical spectrum and highly concentrated laser beams, ISL can currently achieve transmission rates of 10 Gb/s over thousands of kilometers, with advanced multiplexing schemes potentially extending throughput to Tb/s level \cite{gwang2024free}. Consequently, the communication latency between the current and incoming satellites is mainly dominated by the propagation delay over the long inter-satellite distance, which can be reasonably assumed as $T_p$, while the transmission delay is negligible \cite{resswein2026constellation}. This work focuses on the adaptive control of satellite backlog at handover to prevent post-handover task congestion or overly conservative offloading. The study
 on ISL is left for future work.


\subsection{Task Computation and Queue Models}

To efficiently characterize the dynamics of available computing resources and task congestion conditions across the system, we present the task computation and queue models in this subsection. We denote the available computing resources (i.e., the maximum processable number of CPU cycles) of server $i$ at time slot $n$ as $F_i(n), i \in \mathcal{I} = \{u,b,h,s\}$. Let $Q_{v,0}(n)$ signify the local task data backlog on vessel $v$ and $Q_{v,i}(n)$ denote the backlog of vessel $v$ on the server $i$ at the beginning of time slot $n$. They are respectively expressed as
\begin{align*}
Q_{v,0}(n) & = Q_{v,0}(n-1) - \sum_{u \in \mathcal{U}} a_{v,u}(n-1) D_{v,u}^{tx}(n-1) \\ & \hspace{0.5cm} - a_{v,b}(n-1) D_{v,b}^{tx}(n-1), \tag{7a}
\end{align*}
\begin{align*}
Q_{v,i}(n) & = Q_{v,i}(n-1) + D_{v,i}^{tx}(n-1) - D_{v,i}^{comp}(n-1) \\ 
& \hspace{0.5cm} - \boldsymbol{1}_{\{ i \in \{u,h\} \}} \cdot D_{v,i,i'}^{tx}(n-1), i' \in \{h,s\}, \tag{7b}
\end{align*}
where
\vspace{-0.2cm}
\begin{equation*}
\left.D_{v,i}^{tx}(n-1)=\left\{
\begin{array}
{ll}a_{v,b}(n-1) D_{v,b}^{tx}(n-1),  \hspace{0.1cm} i=b, \\
a_{v,u}(n-1) D_{v,u}^{tx}(n-1), \hspace{0.1cm} i=u, \\
\sum_{u\in \mathcal{U}} a_{v,u,h}(n-1) D_{v,u,h}^{tx}(n-1), i=h, \\
a_{v,h,s}(n-1) D_{v,h,s}^{tx}(n-1), \hspace{0.1cm} i=s,
\end{array}\right.\right.
\end{equation*}
and
\vspace{-0.1cm}
\begin{equation*}
\left.D_{v,i,i'}^{tx}(n-1)=\left\{
\begin{array}
{ll} a_{v,u,h}(n-1) D_{v,u,h}^{tx}(n-1), \hspace{0.1cm} i=u, \\
a_{v,h,s}(n-1) D_{v,h,s}^{tx}(n-1), \hspace{0.1cm} i=h.
\end{array}\right.\right.
\end{equation*}
In (7a), $D_{v,u}^{tx}(n-1)$ and $D_{v,b}^{tx}(n-1)$ refer to the task data amount offloaded from vessel $v$ to UAV $u$ and BS in time slot $n-1$, respectively. $a_{v,u}$ and $a_{v,b}$ are binary indicators of vessel $v$'s offloading decision in each time slot. A value of 1 indicates that vessel $v$ chooses the corresponding UAV/BS to offload data. In terms of (7b), $D_{v,i}^{tx}(n-1)$ denotes the total data volume of vessel $v$ received by server $i$, and $D_{v,i}^{comp}(n-1)$ is the volume of vessel $v$'s task data computed by server $i$ in time slot $n-1$. $D_{v,i,i'}^{tx}(n-1)$ signifies the amount of vessel $v$'s data offloaded from UAV $u$ to HAP or HAP to satellite. $a_{v,u,h} = 1$ indicates UAV $u$ chooses to offload vessel $v$'s data to HAP and $a_{v,h,s} = 1$ means HAP offloads vessel $v$'s data to satellite. The actual amount of transmitted data is bounded by link capacity and data backlog at each device. Specifically, $D_{v,i}^{tx}(n) = \min ( R_{v,i}^*(n)\tau, Q_{v,0}(n)), i \in \{b,u\} $ and $D_{v,i,i'}^{tx}(n) = \min ( R_{i,i'}^*(n)\tau, Q_{v,i}(n)), (i,i')\in\{(u,h),(h,s)\}$, where $R_{v,i}^*(n)$ and $R_{i,i'}^*(n)$ are the ultimate rates obtained after all optimization variables are determined.

The energy consumption of UAV $u$ for computing the task of vessel $v$ in time slot $n$ is given by
\begin{equation*}
E_{v,u}^{comp}(n) = \frac{\kappa_u \left(C_v D_{v,u}^{comp}(n)\right)^3}{\tau^2}, \tag{8}
\end{equation*}
where $\kappa_u$ is the energy efficiency factor depending on the chip architecture of UAV $u$'s processor \cite{zlin2024maritime}. Accordingly, the total computation energy consumption of UAV $u$ in the time slot $n$ is given by $E_u^{comp}(n) = \sum_{v \in \mathcal{V}} E_{v,u}^{comp}(n)$.



\subsection{UAV Motion and Energy Model}

\subsubsection{UAV Motion Model}

The displacement of UAV $u$ between two consecutive time slots is given by $\Delta d_{u}(n)=\|\boldsymbol{W_u}(n)-\boldsymbol{W_u}(n-1)\|$. Let $S_{u,max}$ denote the maximum speed of UAV $u$. Accordingly, the UAV motion is subject to the following maximum velocity constraint
\begin{equation*}
\frac{\Delta d_{u}(n)}{\tau}\leq S_{u,max},\forall u, n. \tag{9}
\end{equation*}
To avoid collisions among UAVs, the inter-UAV distances must be maintained above the minimum safety distance denoted by $d_{safe}$. Hence, the positions of any pair of UAV $u$ and $u'$ must comply with the following constraint:
\begin{equation*}
\| \boldsymbol{W_u}(n)-\boldsymbol{W_{u'}}(n)\|\geq d_{safe}, \forall u,u' \in \mathcal{U}, u\neq u', \forall n. \tag{10}
\end{equation*}

\subsubsection{UAV Energy Model}
The average velocity of the UAV $u$ between time slots $n-1$ and $n$ can be expressed as $\overline{s}_u(n) = \frac{\Delta d_{u}(n)}{\tau}$, thus the corresponding flight energy consumption \cite{sjung2023marine} is $E_u^{flight}(n) = 0.5M_u \tau {\overline{s}_u(n)}^2$, where $M_u$ is the mass of UAV $u$. To address the difficulties of UAV charging and emergency landing in maritime environments while ensuring service continuity, each UAV is required to reserve sufficient residual energy for two essential operations: uploading its remaining local task backlog to HAP for service handover and returning to the charging station. Based on the above derivation, the energy required for UAV $u$ to return to the charging station at the beginning of time slot $n$ can be expressed as
\begin{equation*}
E_u^{cs}(n)=0.5 M_u \| \boldsymbol{W_u}(n-1)-\boldsymbol{W_{cs}}\| S_{u,auto}, \tag{11}
\end{equation*}
where $\boldsymbol{W_u}(n-1)$ denotes its final position in the previous time slot, which is also its initial position at time slot $n$, and $S_{u,auto}$ is the autonomous return flight speed. The energy required for UAV $u$ to transmit the remaining data to HAP is given by
\begin{equation*}
E_u^{re}(n)=\frac{\sum_{v\in\mathcal{V}}Q_{v,u}(n)}{R_{u,h}(n)}\cdot P_u(n). \tag{12}
\end{equation*}
The residual energy of UAV $u$ at the start of time slot $n$ is
\begin{equation*}
E_u(n) = E_u(n-1) - E_u^{cons}(n-1), \tag{13}
\end{equation*}
where $E_u^{cons}(n) = E_u^{tx}(n) + E_u^{comp}(n) + E_u^{flight}(n)$ is the total energy consumption of UAV $u$ in time slot $n$.

Based on above establishment, we represent the total execution delay of vessel $v$'s task as
\begin{equation*}
T_v = (N_v + 1) \cdot \tau , \tag{14}
\end{equation*}
where $N_v$ satisfies the following two formulas simultaneously
\begin{equation*}
Q_{v,0}(N_v) + \sum_{i \in \mathcal{I}} Q_{v,i}(N_v) > 0, \tag{15a}
\end{equation*}
\begin{equation*}
Q_{v,0}(N_v + 1) + \sum_{i\in \mathcal{I}} Q_{v,i}(N_v + 1) = 0. \tag{15b}
\end{equation*}
The above formulas indicate that its task data is just cleared from all queues in the system at the beginning of time slot $N_v + 1$. Since the time slot index starts from 0, the number of time slots required to finish the task of vessel $v$ is $N_v + 1$.


\section{Problem Formulation}

In this section, we first formulate the overall problem based on the system established above. Subsequently, we provide a detailed problem description and introduce the overarching decomposition framework to facilitate problem-solving.

Our overall objective is to minimize the total task execution delay of all vessels in the system by jointly optimizing the multi-layer task offloading decisions, bandwidth allocation policies of BS and UAVs, computing resource allocation across heterogeneous servers, and UAV trajectory planning. The overall problem can be mathematically formulated as

$\mathcal{P}_0$: \textbf{Overall Problem}
\begin{align*}
\min _{\substack {\boldsymbol{a}, \boldsymbol{B}, \\ \boldsymbol{D^{comp}}, \boldsymbol{W_U}}}
& \sum_{v \in \mathcal{V}} T_v\\
\text {s.t. } \hspace{0.3cm} &a_{v,u}(n), a_{v,b}(n), a_{v,u,h}(n), a_{v,h,s}(n) \in \{0,1\}, \\
&\forall v,u,n, \tag{16a}\\
& \sum_{u \in \mathcal{U}} a_{v,u}(n) + a_{v,b}(n) = 1, \forall v,n, \tag{16b}\\
& B_{v,i}(n) \geq 0, \forall i \in \{u,b\},v,n, \tag{16c}\\
& \sum_{v \in \mathcal{V}} a_{v,i}(n) B_{v,i}(n) \hspace{-0.6mm} \leq \hspace{-0.6mm} B_i(n), \forall i \in \{u,b\}, n, \hspace{-0.2cm} \tag{16d}\\
& \sum_{v \in \mathcal{V}} D_{v,i}^{comp}(n) C_{v} \leq F_i(n), \forall i,n, \tag{16e}\\
& D_{v,i}^{comp}(n) \leq Q_{v,i}(n) + D_{v,i}^{tx}(n), \\ 
& \forall v, i \in \{b,s\}, n, \tag{16f} \\
& D_{v,i}^{comp}(n) \leq Q_{v,i}(n) + D_{v,i}^{tx}(n) - D_{v,i,i'}^{tx}(n), \\
& \forall v, (i, i') \in \{(u, h), (h,s)\}, n, \tag{16g} \\
& D_{v,i}^{comp}(n)\geq0,\forall v,i,n, \tag{16h} \\
& E_u(n+1) \geq E_u^{cs}(n+1) + E_u^{re}(n+1), \forall u,n, \tag{16i} \\
& \text{(9) and (10)}.
\end{align*}

For the optimization variables, the offloading decision set $\boldsymbol{a} = \{ \boldsymbol{a_{v,u}}(n), a_{v,b}(n), \boldsymbol{a_{v,u,h}}(n), \boldsymbol{a_{v,h,s}}(n) | \forall v,u,n \}$ contains the decision variables of each vessel, UAV, and the HAP in each time slot, where $\boldsymbol{a_{v,u}}(n) = [ a_{v,1}(n), \dots, a_{v,U}(n) ]^T$ and $a_{v,b}(n)$ denote the V2U and V2B offloading decision of each vessel $v$, respectively, $\boldsymbol{a_{v,u,h}}(n) = [ a_{1,u,h}(n), \dots, a_{V,u,h}(n) ]^T$ signifies U2H offloading decision of each UAV $u$, and $\boldsymbol{a_{v,h,s}}(n) = [ a_{1,h,s}(n), \dots, a_{V,h,s}(n) ]^T$ is the H2S offloading decision of the HAP. $\boldsymbol{B} = \{ \boldsymbol{B_u}(n), \boldsymbol{B_b}(n) | \forall u,n\}$ includes the bandwidth allocation policy of each UAV $u$ and the BS, where $\boldsymbol{B_u}(n) = \{ B_{v,u}(n) | \forall v \}$ and $\boldsymbol{B_b}(n) = \{ B_{v,b}(n) | \forall v \}$. $\boldsymbol{D^{comp}} = \{ D_{v,i}^{comp}(n) | \forall v,i,n\}$ consists of the computing resource allocation of each server. $\boldsymbol{W_U} = \{ \boldsymbol{W_u}(n) | \forall u,n\} $ comprises the trajectory of each UAV.

In terms of constraints, (16a) defines the offloading decision indicators of vessels and servers as binary variables, while (16b) ensure that each vessel offloads data to one UAV or BS in each time slot. (16c) imposes the non-negativity of bandwidth allocated to each vessel. (16d) guarantees that the total bandwidth allocated by each UAV or the BS does not exceed its real-time available spectrum. (16e) ensures the total allocated computing resources of each server do not exceed its available capacity. (16f) guarantees that the amount of vessel $v$'s data computed by the BS or satellite in each time slot is bounded by its available backlog after the server receiving its newly offloaded data. (16g) applies similar bounds to UAVs and the HAP, while further excluding the data forwarded to upper-layer servers. (16h) imposes the non-negativity of the computed data volume. (16i) ensures that, after completing the operations in each time slot, the residual energy of each UAV at the beginning of next time slot can support both backlog handover and returning to the charging station. (9) and (10) are the UAV maximum velocity and collision-avoidance constraints, respectively.

The problem $\mathcal{P}_0$ is an intricate optimization problem arising from the highly dynamic and deeply hierarchical SAGSIN. Specifically, the heterogeneous network architecture introduces coupled task offloading decisions among four layers, while the coexistence of bandwidth, task computation, UAV mobility, and residual energy constraints further leads to strongly intertwined optimization variables. Consequently, obtaining the globally optimal solution of this problem in polynomial time is intractable. To tackle this challenge, we first develop a low-complexity task offloading algorithm tailored to SAGSIN, which substantially reduces the layer-wise decision-making complexity while adapting to real-time multifaceted system states. Thereafter, UAV-BS bandwidth allocation, UAV trajectories, and computing resource allocation are jointly refined via a block coordinate descent (BCD) framework to efficiently handle the coupled resource optimization.



\section{Dynamic Task Offloading Algorithm for SAGSIN}

The task offloading in SAGSIN involves twofold challenges: the abrupt shift in satellite states caused by handover and the high complexity of task scheduling over the dynamic heterogeneous network. To tackle these issues, this section first introduces the anticipatory strategy for handling satellite handover and then elaborates on the dynamic task scheduling algorithm for SAGSIN.



\subsection{Anticipatory Satellite Handover Strategy}

The satellite offloading strategy designed solely based on the current satellite state can be myopic. If the incoming satellite possesses significantly lower computing capability, aggressive offloading before handover will lead to task congestion afterwards. Conversely, if it offers substantially stronger computing capability, such offloading strategy may become conservative. Therefore, an adaptive offloading strategy accounting for both satellites is essential to smooth out such abrupt changes.

\begin{figure}[!t]
\centering
\includegraphics[width=3.0 in]{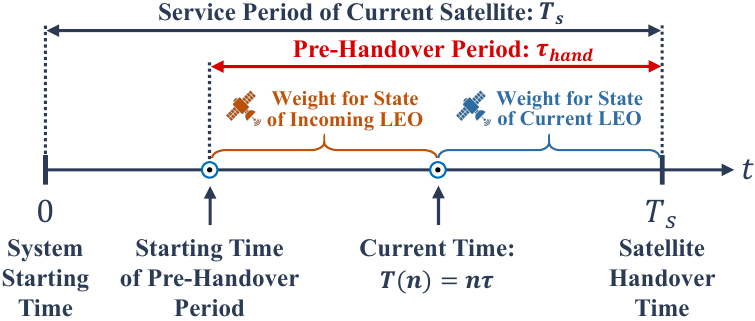}
\caption{Illustration of the real-time joint satellite state for anticipatory handover strategy.}
\label{handover}
\end{figure}

To this end, we formulate the real-time joint states of the current and incoming satellites to guide offloading decision-making, adaptively regulating the amount of data offloaded to the satellite in advance of handover. A pre-handover period $\tau_{hand}$ is set prior to the formal handover time $T_s$, as illustrated in Fig.~\ref{handover}. Once the system time $T(n)$ enters this period, the real-time joint states of available computing resources and task backlog for both satellites are respectively defined as
\begin{equation*}
\widetilde{F}_s(n) = \frac{T_s - T(n)}{\tau_{hand}} F_s(n) + \left(1-\frac{T_s - T(n)}{\tau_{hand}}\right) F_{s'}(n),  \tag{17a}
\end{equation*}
\begin{equation*}
\widetilde{Q}_{v,s}(n) = \frac{T_s - T(n)}{\tau_{hand}} Q_{v,s}(n) + \left(1-\frac{T_s - T(n)}{\tau_{hand}}\right) Q_{v,s'}(n), \tag{17b} 
\end{equation*}
where $F_{s'}(n)$ and $Q_{v,s'}(n)$ denote the available computing resources and vessel $v$'s backlog at time slot $n$ on the incoming satellite. Since it has not started serving this region yet, its backlog can be initialized as zero or flexibly configured based on practical demands. During the pre-handover period, the real-time joint states of satellites are dynamic weighted sums of states for the current and incoming satellites, where the weight of the incoming one increases as handover approaches. These joint states are employed to guide the task offloading strategy, thereby smoothing the transition and enhancing satellite resource utilization. The duration of $\tau_{hand}$ is designed as
\begin{equation*}
\tau_{hand} = \min \left\{ T_s, \left\lceil \xi \cdot \frac{|\overline{F}_s - \overline{F}_{s'}|} {\overline{C}_v \cdot \overline{D}_v} \right\rceil\cdot\tau\right\}, \tag{18}
\end{equation*}
where $\overline{F}_s$ and $\overline{F}_{s'}$ denote average computing resources of the two satellites, $\overline{C}_v$ and $\overline{D}_v$ are average computational density and data size among vessels, respectively, and $\xi$ is a tuning parameter. It indicates that a larger capability gap between the two satellites requires a longer preparation period and the system should immediately enter the pre-handover stage when the remaining service time is insufficient.


\subsection{Dynamic Layer-Wise Task Scheduling for SAGSIN}

The multi-layer architecture of SAGSIN leads to prohibitive complexity in optimizing task offloading decisions. To address this challenge, we develop a low-complexity approach inspired by backpressure (BP) routing theory to decompose the combinatorial problem into elegant layer-wise task data scheduling, thereby reducing complexity while adapting to multi-dimensional system dynamics.


\begin{figure}[!t]
\centering
\includegraphics[width=3.4 in]{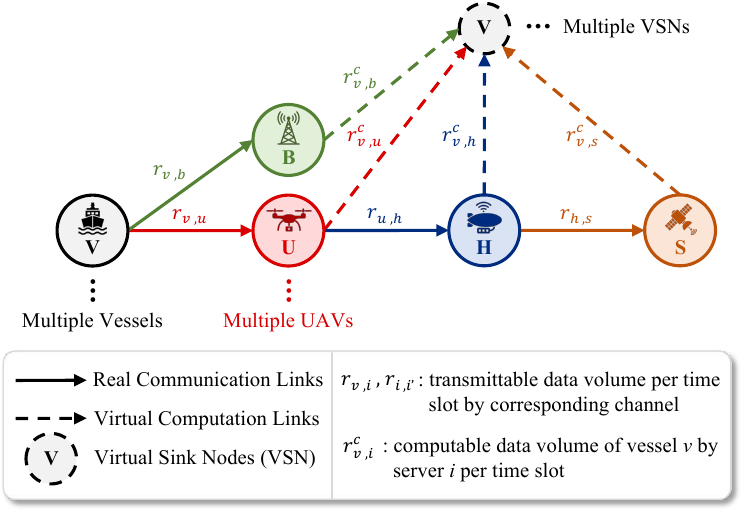}
\caption{The graph model of the proposed SAGSIN architecture containing multiple vessels, multiple UAVs, 1 BS, 1 HAP, and 1 LEO satellite, represented by node V, U, B, H, and S, respectively.}
\label{graph}
\end{figure}

The BP algorithm is a routing scheme for multi-hop networks that determines data forwarding paths according to congestion differentials among network nodes to maximize network throughput. In recent years, it has been widely applied to various fields, such as satellite networks, mobile ad hoc networks (MANETs), and traffic signal control \cite{xdeng2023distance}, \cite{she2025back}. Unlike the classical BP algorithm, the destination server of each vessel’s task data should be adaptively optimized rather than predetermined. To align with routing theory, the network is remodeled into a graph as illustrated in Fig.~\ref{graph}. A virtual sink node (VSN) is established for each vessel $v$. Each server node is connected to all VSNs via virtual computation links. We first homogenize the communication and computing capabilities, laying the foundation for designing the informative task scheduling indicators. We define the rate of each real communication link (i.e., $r_{v,u}$, $r_{v,b}$, $r_{u,h}$, and $r_{h,s}$) as the corresponding channel capacity $R$ times $\tau$, and the rate of virtual computation link between server $i$ and VSN $v$ as $r_{v,i}^c = F_i / C_v$. The time-slot index $n$ is omitted in Fig.~\ref{graph}. In this way, the rate $r_e$ of each link $e$ is unified as the number of bits that the corresponding channel can transmit or the server can compute within one time slot, while the virtual destination of each vessel’s data is specified as its VSN. The distance of each link $e$ is defined as $w_e = \overline{r} \cdot r_{max} / r_e$, where $\overline{r}$ and $r_{max}$ are the average and maximum link rates in the system.

To guide task data towards servers with low congestion and abundant communication and computing resources, we define the Pressure Index (PI) of vessel $v$ and server $i$ with respect to vessel $v$'s data at time slot $n$ respectively as
\begin{equation*}
\left\{
\begin{aligned}
J_{v,0}(n) &= Q_{v,0}(n) + w_{v,0}^{min}(n), \\
J_{v,i}(n) &= Q_{v,i}(n) + w_{v,i}^{min}(n), \\
\end{aligned}
\right. \tag{19}
\end{equation*}
where $w_{v,0}^{min}(n)$ and $w_{v,i}^{min}(n)$ denote the shortest distances from that node to VSN $v$, which can be easily obtained by comparing the distances of several paths. The PI functions as an informative indicator that integrates multifaceted real-time attributes of each device regarding vessel $v$, including task congestion level, available computing capability, rate of links to high-capacity servers, and network topology. A smaller PI reflects a node with a lighter task load, stronger local computing capability, or faster access to high-capacity servers. As the driving force for task offloading, the Pressure Differentials (PDs) of vessel $v$'s task data on the links between vessel $v$ and UAV $u$ or the BS, between UAV $u$ and HAP, and between HAP and satellite are respectively given by
\begin{equation*}
\left\{
\begin{aligned}
& \Delta J_{v \to i}(n) = J_{v,0}(n) - J_{v,i}(n), \forall i \in \{u,b\}, \\
& \Delta J_{v,u \to h}(n) = \max\{J_{v,u}(n) - J_{v,h}(n), 0\}, \\
& \Delta J_{v,h \to s}(n) = \max\{J_{v,h}(n) - J_{v,s}(n), 0 \}.
\end{aligned}
\right. \tag{20}
\end{equation*}
Then the offloading decision of each vessel $v$ is the solution to the following formulation subject to constraint (16b)
\begin{equation*}
\boldsymbol{a_{v,i}}(n)^* = \underset{ i \in \{u, b\}} {\operatorname{argmax}}\boldsymbol{a_{v,i}} (n)^T \left[\Delta\boldsymbol{J_{v \to i}}(n) \odot \boldsymbol{r_{v,i}}(n) \right], \tag{21}
\end{equation*}
where the vector $\boldsymbol{a_{v,i}}(n) = [a_{v,1}(n), \dots, a_{v,U}(n), a_{v,b}(n)]^T$, $\Delta\boldsymbol{J_{v \to i}}(n) = [ \Delta J_{v \to 1}(n), \dots, \Delta J_{v \to U}(n), \Delta J_{v \to b}(n) ]^T$, and $\boldsymbol{r_{v,i}}(n) = [ r_{v,1}(n), \dots, r_{v,U}(n), r_{v,b}(n) ]^T$. This mechanism leads each vessel to offload its task data to the UAV or BS offering the best combination of low load and high throughput, thereby avoiding server overload while fully exploiting high-rate channels. The offloading decision of each UAV $u$ and the HAP is expressed as
\begin{equation*}
\boldsymbol{a_{v,i,i'}}(n)^* = \underset{ v \in \mathcal{V}} {\operatorname{argmax}}\boldsymbol{a_{v,i,i'}}(n)^T \Delta\boldsymbol{J_{v, i \to i'}}(n), \tag{22}
\end{equation*}
where $\Delta\boldsymbol{J_{v, i \to i'}}(n) = [ \Delta J_{1, i \to i'}(n), \dots, \Delta J_{V, i \to i'}(n) ]^T$ and $\boldsymbol{a_{v,i,i'}}(n) = [ a_{1,i,i'}(n), \dots, a_{V,i,i'}(n) ]^T$ subject to $\sum_{v \in \mathcal{V}} a_{v,i,i'}(n) \leq 1$, $(i,i') \in \{(u,h), (h,s)\}$. This indicates that each UAV and the HAP selects a vessel with the most urgent offloading demands (e.g., severe backlog or the possibility of fast access to the upper-layer server with more computing resources) and transmits its data to HAP and satellite, respectively. Selecting only one vessel's data prevents servers with low backlog from indiscriminately forwarding excessive buffered data to upper-layer servers, thereby avoiding potential load imbalance. The above task offloading approach is summarized in \textbf{Algorithm~\ref{alg:offloading}} for clarity.

By shifting from global coordination to layer-wise pressure-driven forwarding, our scheme reduces the complexity from exponential level to approximately $\mathcal{O}(VU)$. Furthermore, since the PI is updated at each time slot based on instantaneous system states, it possesses strong adaptability to the high dynamics of SAGSIN environments.

\begin{algorithm}[!t]
    \caption{Dynamic Task Offloading for SAGSIN.}
    \label{alg:offloading}
    \renewcommand{\algorithmicrequire}{\textbf{Input:}}
    \renewcommand{\algorithmicensure}{\textbf{Output:}}
    
    \begin{algorithmic}[1]
        \REQUIRE $\mathcal{V}$, $\mathcal{I}$, $D_v$, $C_v$, optimized $\boldsymbol{W_u}$ at previous time slot, and all other relevant parameters at current time slot.   
        \ENSURE V2U/B offloading decisions $\boldsymbol{a_{v,i}}(n)^*, \forall v$, U2H offloading decisions $\boldsymbol{a_{v,u,h}}(n)^*,\forall u$, and H2S offloading decision $\boldsymbol{a_{v,h,s}}(n)^*$ at current time slot $n$.    
        
        \STATE Initialize the network graph. Calculate the rates $r_{v,u}$, $r_{v,b}$, $r_{u,h}$, $r_{h,s}$ for all real communication links and the rates $r_{v,i}^c$ for all virtual computation links.

        \STATE Calculate the distance of each link $e$ by $w_e = \overline{r} \cdot r_{max} / r_e$.
        
        \STATE Calculate the PI of all nodes and PD of all real links using (19) and (20), respectively.

        \FOR{$v \in \mathcal{V}$}
            \STATE Solve (21) by selecting the UAV or BS corresponding to the link with maximum product of PD and $r_{v,i}$ to obtain the offloading decision $\boldsymbol{a_{v,i}}(n)^*$ of vessel $v$.
        \ENDFOR

        \FOR{$u \in \mathcal{U}$}
            \STATE Solve (22) by selecting the vessel user that has maximum PD on this U2H link, thereby determining the offloading decision $\boldsymbol{a_{v,u,h}}(n)^*$ of UAV $u$.
        \ENDFOR

        \STATE Solve (22) by selecting the vessel user that has maximum PD on this H2S link to obtain the offloading decision $\boldsymbol{a_{v,h,s}}(n)^*$ of the HAP.

    \end{algorithmic}
\end{algorithm}


\section{Joint UAV-BS Bandwidth, UAV Trajectory, and Computing Resource Optimization}

Building upon the macroscopic multi-layer task offloading strategy derived according to diverse system states, this section focuses on the fine-grained orchestration of underlying resources to further enhance the system performance.

Specifically, we jointly optimize the UAV-BS bandwidth allocation, UAV trajectory planning, and computing resource allocation of servers. The computational complexity of the proposed overall method is then analyzed. Due to the presence of inter-coupled multi-dimensional variables, this problem is highly intricate. Motivated by the principle of the block coordinate descent (BCD) method \cite{jzhang2026energy}, which is well adapted to high-dimensional optimization, we decompose the joint optimization problem into three tractable subproblems to reduce the complexity. Tailored solutions are developed according to the characteristics of each subproblem and they are iteratively optimized until converging to an efficient overall solution.

\subsection{Bandwidth Allocation Policy of UAVs and BS}

This subproblem aims to maximize the total transmission rate of communication links between each UAV/BS and the vessels it serves by optimizing bandwidth allocation, thereby boosting the volume of offloaded vessel data and reducing overall task delay. This subproblem can be formulated as

$\mathcal{P}_1$: \textbf{UAV and BS Bandwidth Allocation}
\begin{align*}
\max _{\substack {\boldsymbol{B_u}(n),\\ \boldsymbol{B_b}(n)}}
& \sum_{i \in \{u,b\}} \sum_{v \in \mathcal{V}} a_{v,i}(n) B_{v,i}(n) \log_2 \left( 1 + \frac{P_v(n) h_{v,i}(n) } {B_{v,i}(n) N_0} \right) \\
\text {s.t. } \hspace{0.1cm} & \hspace{0.3cm} \text{(16c), (16d), and (16i)},
\end{align*}
With determined vessel offloading decisions, the second-order derivative of the objective function w.r.t. $B_{v,i}(n)$ is $-\frac{{P_v(n)}^2 {h_{v,i}(n)}^2}{\left[B_{v,i}(n) N_0 + P_v(n) h_{v,i}(n) \right]^2 B_{v,i}(n) \ln2} < 0$, proving that it is strictly concave in $B_{v,i}(n)$. For the constraints, (16c) and (16d) are affine and also convex with $B_{v,i}(n)$. (16i) can be rewritten as $\sum_{v \in \mathcal{V}_u(n)} B_{v,u}(n) \log_2\left(1 + \frac{P_v(n) h_{v,u}(n)} { B_{v,u}(n) N_0}\right) \leq A_u^1(n), \forall u$, where $\mathcal{V}_u (n)$ is the set of vessels served by UAV $u$ at time slot $n$, $A_u^1(n) = \frac{[E_u(n+1) - E_u^{cs}(n+1) - A_u^2(n)] R_{u,h}(n+1)}{P_u(n+1) \tau}$, and $A_u^2(n) = \frac{\sum_{v \in \mathcal{V}} Q_{v,u}(n) - D_{v,u}^{comp}(n) - a_{v,u,h}(n) D_{v,u,h}^{tx}(n)} {R_{u,h}(n+1)} P_u(n+1)$. It is strictly concave in $B_{v,i}(n)$, by similar reasoning as the objective function. Thus, $\mathcal{P}_1$ is a convex optimization problem.

\textbf{Proposition 1:} The optimal bandwidth allocation policies of the coastal BS and UAVs are expressed as
\begin{equation*}
\left\{
\begin{aligned}
& B_{v,b}^*(n) = \frac{ P_v(n) h_{v,b}(n)} {\sum_{v \in \mathcal{V}_b(n)} P_v(n) h_{v,b}(n)} B_b(n), v \in \mathcal{V}_b(n),\\
& B_{v,u}^*(n) = \frac{ P_v(n) h_{v,u}(n)} {\sum_{v \in \mathcal{V}_u(n)} P_v(n) h_{v,u}(n)} \overline{B_u}(n), v \in \mathcal{V}_u(n), \\
\end{aligned} \tag{23}
\right.
\end{equation*}
where $\overline{B_u}(n) = \min \{B_u(n), B_u'(n)\}$ and $B_u'(n)$ satisfies the equation $B_u'(n) \log_2\left(1 + \frac{\sum_{v \in \mathcal{V}_u(n)} P_v(n) h_{v,u}(n)} {B_u'(n) N_0}\right) = A_u^1(n)$. Since the left-hand side (LHS) of above equation is monotonically increasing with $B_u'(n)$, the solution can be efficiently obtained via the bisection method.

\emph{Proof:} For the bandwidth allocation of BS, the Lagrange function is expressed as
\begin{align*}
L(B_{v,b}(n), \lambda_b) =& \sum_{v \in \mathcal{V}_b(n)} B_{v,b}(n) \log_2 \left(1 + \frac{P_v(n) h_{v,b}(n)}{B_{v,b}(n) N_0 } \right) \\ 
& + \lambda_b \left(\sum_{v \in \mathcal{V}_b(n)} B_{v,b}(n) - B_b(n) \right), \tag{24}
\end{align*}
where $\lambda_b$ is the Lagrange multiplier associated with constraint (16d). According to KKT conditions \cite{wwu2025multi}, the optimal $B_{v,b}(n)$ is obtained when $\frac{\partial L}{\partial B_{v,b}(n)} = 0$, i.e.,
\begin{equation*}
\log_2 \left(1 + \frac{\gamma_{v,b}(n)} {B_{v,b}(n)}\right) - \frac{\gamma_{v,b}(n)}{[B_{v,b}(n) + \gamma_{v,b}(n)]\ln2}+\lambda_b=0, \tag{25}
\end{equation*}
where $\gamma_{v,b}(n) = \frac{P_v(n) h_{v,b}(n)}{N_0}$. Let $\frac{\gamma_{v,b}(n)}{B_{v,b}(n)}=m_v$, (25) can be rewritten as $\log_2 (1 + m_v) - \frac{m_v}{(1 + m_v)\ln2} = - \lambda_b$. Since the LHS of this equation increases monotonically with $m_v$, the solution for $m_v$ is unique. Hence, we have $\frac{\gamma_{v,b}(n)}{B_{v,b}(n)} = \beta, \forall v \in \mathcal{V}_b(n)$, where $\mathcal{V}_b(n)$ is the vessel set served by BS. The optimal solution occurs only when (16d) is active, as allocating any remaining bandwidth will further increase the data rate. Then we have $\sum_{v \in \mathcal{V}_b(n)} B_{v,b}^*(n) = \sum_{v \in \mathcal{V}_b(n)} \frac{\gamma_{v,b}(n)}{\beta^*} = B_b(n)$. Thus, $\beta^* = \sum_{v \in \mathcal{V}_b(n)} \gamma_{v,b}(n)/B_b(n)$. Substituting $\beta^*$ into $B_{v,b}(n) = \frac{\gamma_{v,b}(n)}{\beta}$, we obtain the optimal solution.

As for UAV bandwidth allocation, following a derivation similar to the above, the solution satisfying (16c) and (16d) is $B_{v,u}(n) = \frac{ P_v(n) h_{v,u}(n)} {\sum_{v \in \mathcal{V}_u(n)} P_v(n) h_{v,u}(n)} B_u(n)$. If it satisfies (16i), it will be the final optimal solution. Otherwise, if (16i) is violated, the maximum allocable bandwidth should be reduced to $B_u'(n)$ as described in \textbf{Proposition 1}, where the total data rate under optimal allocation is equal to $A_u^1(n)$, thereby ensuring (16i) is satisfied. This completes the proof.
$\hfill\blacksquare$


\subsection{UAV Trajectory Optimization}


Beyond bandwidth resource dimension, the mobilities of UAVs provide a crucial degree of freedom to dynamically reconstruct the network topology. By optimizing their trajectories, we directly modulate the V2U channel states. Specifically, the trajectory planning aims to strategically position UAVs to maximize the aggregate data rate and throughput for associated vessels while strictly adhering to kinematic, safety, and residual energy constraints. Retaining the relevant terms, this subproblem is accordingly formulated as

$\mathcal{P}_2$: \textbf{UAV Trajectory Optimization}
\begin{align*}
\max _{\substack {\boldsymbol{W_u}(n)}}
& \sum_{u \in \mathcal{U}} \sum_{v \in \mathcal{V}} a_{v,u}(n) B_{v,u}(n) \log_2 \left( 1 + \frac{P_v(n) h_{v,u}(n) } {B_{v,u}(n) N_0} \right)\\
\text {s.t. } & \text{(9), (10), and (16i)}.
\end{align*}

The non-convexity of the objective function, constraint (10) and (16i) poses the primary challenge. Further analysis reveals that both the objective function to be maximized and the LHS of constraint (10), which is required to exceed a threshold, can be effectively relaxed using convex lower bounds. A similar strategy can also be applied to handle constraint (16i). Motivated by this insight, we employ the successive convex approximation (SCA) technique to convert the original problem into a series of approximated convex problems, which are solved iteratively until convergence.

\textbf{Proposition 2:} The problem $\mathcal{P}_2$ can be converted into the approximated convex problem expressed as

$\mathcal{P}_2'$: \textbf{Convex Approximation Problem of $\mathcal{P}_2$}
\begin{align*}
\max _{\substack {\boldsymbol{W_u}(n),\\ z_{v,u}(n)}}
& \sum_{u \in \mathcal{U}} \sum_{v \in \mathcal{V}_u(n)} \widetilde{R}_{v,u}(n)\\
\text {s.t. } & \text{(9)}, \\
& {\widetilde{d}_{u,u'}(n)}^2 \geq {d_{safe}}^2, \forall u,u' \in \mathcal{U}, u\neq u', \tag{26a}\\
& E_u^{flight}(n) + E_u^{cs}(n+1) + A_u^3(n+1) \tau \cdot \\
& \sum_{v \in \mathcal{V}_u(n)} B_{v,u}(n) \log_2 \left(1 + \frac{\gamma_{v,u}(n)} {z_{v,u}(n)}\right) \leq A_u^4(n), \forall u, \tag{26b} \\
& z_{v,u}(n) \leq {H_u}^2 + \|\boldsymbol{W_u^{(k)}}(n)-\boldsymbol{W_v}(n)\|^{2} \\
&+ 2\left(\boldsymbol{W_u^{(k)}}(n)-\boldsymbol{W_v}(n)\right)^{T} \left(\boldsymbol{W_u}(n) - \boldsymbol{W_u^{(k)}}(n)\right), \\
&\forall v,u, \tag{26c}
\end{align*}
where $\widetilde{R}_{v,u}(n)$ and ${\widetilde{d}_{u,u'}(n)}^2$ represent the convex lower bound of $R_{v,u}(n)$ and $\| \boldsymbol{W_u}(n)-\boldsymbol{W_{u'}}(n)\|^2$, respectively, $z_{v,u}(n)$ is the introduced auxiliary variable, $\boldsymbol{W_u^{(k)}}(n)$ denotes the solution of UAV $u$'s position at the $k$-th iteration, $A_u^3(n+1) = \frac{P_u(n+1)}{R_{u,h}(n+1)}$, and $A_u^4(n) = E_u(n) - E_u^{tx}(n) - E_u^{comp}(n) -\frac{\sum_{v \in \mathcal{V}} Q_{v,u}(n) - D_{v,u}^{comp}(n) - a_{v,u,h}(n) D_{v,u,h}^{tx}(n)} {R_{u,h}(n+1)} \cdot P_u(n+1)$. $\mathcal{P}_2'$ can be efficiently solved by CVX solver. In each iteration, it is constructed based on the solution of previous iteration and solved iteratively until converging to a near-optimal solution.


\emph{Proof:} In the objective function, $R_{v,u}(n)$ is a non-convex function w.r.t. the variable $\boldsymbol{W_u}(n)$, but it is convex with the entire $\| \boldsymbol{W_u}(n) - \boldsymbol{W_v}(n)\|^2$. Let $\phi_{v,u}(n) = \| \boldsymbol{W_u}(n) - \boldsymbol{W_v}(n)\|^2$, $R_{v,u}(n)$ can be globally lower-bounded by its first-order Taylor expansion with $\phi_{v,u} (n)$ at any point. Then the lower bound of $R_{v,u}(n)$ is given by
\begin{equation*}
\widetilde{R}_{v,u}(n)=R_{v,u}^{(k)}(n)+\nabla R_{v,u}^{(k)}(n) (\phi_{v,u}(n)-\phi_{v,u}^{(k)}(n)), \tag{27}
\end{equation*}
where $R_{v,u}^{(k)}(n)$ and $\nabla R_{v,u}^{(k)}(n)$ are the data rate between vessel $v$ and UAV $u$ and the first-order derivative of $R_{v,u}(n)$ w.r.t. $\phi_{v,u} (n)$ at the $k$-th iteration, respectively. Specifically,
\begin{align*}
& \nabla R_{v,u}^{(k)}(n) \hspace{-0.6mm}=\hspace{-0.6mm} \frac{- B_{v,u}(n) \gamma_{v,u}(n) \log_2e}{\left({H_u}^2 + \phi_{v,u}^{(k)}(n)\right) \hspace{-1.2mm} \left({H_u}^2 + \phi_{v,u}^{(k)}(n) + \gamma_{v,u}(n)\right)}, \tag{28a}\\
& \hspace{0.8cm} R_{v,u}^{(k)}(n) = B_{v,u}(n) \log_2 \left(1+\frac{\gamma_{v,u}(n)}{ {H_u}^2 + 
\phi_{v,u}^{(k)}(n) }\right), \tag{28b}
\end{align*}
where $\phi_{v,u}^{(k)}(n) = \| \boldsymbol{W_u^{(k)}}(n) - \boldsymbol{W_v}(n) \|^2$ and we let $\gamma_{v,u}(n) = \frac{P_v(n) L_0 |h_{v,u}^R(n)|^2}{B_{v,u}(n) N_0}$. Consequently, we obtain convex lower bound for the objective function.

We rewrite the constraint (10) as $\| \boldsymbol{W_u}(n)-\boldsymbol{W_{u'}}(n)\|^2 \geq d_{safe}^2$. The LHS is non-convex with $\boldsymbol{W_u}(n)$ and $\boldsymbol{W_{u'}}(n)$, but convex with $\boldsymbol{W_u}(n)-\boldsymbol{W_{u'}}(n)$, thus its lower bound can be obtained by utilizing its first-order Taylor expansion at any given point $ \psi_{u,u'}^{(k)}= \boldsymbol{W_u^{(k)}}(n)-\boldsymbol{W_{u'}^{(k)}}(n)$ as
\begin{equation*}
{\widetilde{d}_{u,u'}(n)}^2 \hspace{-0.6mm}=\hspace{-0.8mm} \|\psi_{u,u'}^{(k)}\|^{2} \hspace{-0.5mm}+\hspace{-0.5mm} 2 \psi_{u,u'}^{(k) \hspace{1.5mm} T} \hspace{-0.3mm} ((\boldsymbol{W_u}(n)-\boldsymbol{W_{u'}}(n))-\psi_{u,u'}^{(k)}). \tag{29}
\end{equation*}

(16i) can be rewritten as $E_u^{flight}(n) + E_u^{cs}(n+1) + A_u^3(n+1) \tau \cdot \sum_{v \in \mathcal{V}_u(n)} B_{v,u}(n) \log_2 \left(1 + \frac{\gamma_{v,u}(n)} {{H_u}^2 + \| \boldsymbol{W_u}(n) - \boldsymbol{W_v}(n) \|^2}\right) \leq A_u^4(n)$. In the above inequality, the third term of LHS are non-convex w.r.t. $\boldsymbol{W_u}(n)$. To derive a convex upper bound for the LHS, we introduce the auxiliary variables $z_{v,u}(n)$ satisfying
\begin{equation*}
z_{v,u}(n) \leq {H_u}^2+\| \boldsymbol{W_u}(n) - \boldsymbol{W_v}(n)\|^2 . \tag{30}
\end{equation*}
The feasible region of (30) is still non-convex. To cope with this, a lower bound for convex RHS of (30) is obtained via its first-order Taylor expansion at point $\boldsymbol{W_u^{(k)}}(n)$ as expressed by (26c). Then (26b) is convex with $z_{v,u}(n)$ and (26c) is convex with $z_{v,u}(n)$ and $\boldsymbol{W_u}(n)$. After the above transformation, the objective function and all constraints of $\mathcal{P}_2'$ are convex with $\boldsymbol{W_u}(n)$ and $z_{v,u}(n)$. This completes the proof.
$\hfill\blacksquare$


\subsection{Computing Resource Allocation}

\begin{algorithm}[!t]
    \caption{Two-Stage Computing Resource Allocation.}
    \label{alg:computing}
    \renewcommand{\algorithmicrequire}{\textbf{Input:}}
    \renewcommand{\algorithmicensure}{\textbf{Output:}}
    
    \begin{algorithmic}[1]
        \REQUIRE $\mathcal{V}$, $Q_{v,i}(n)$, $F_i(n)$, $r_{v,i}^c(n)$, $\kappa_u$, $E_u(n)$, $D_v$, and $C_v$.   
        \ENSURE Computing resource allocation strategy $\boldsymbol{F_{v,i}}(n) = \{F_{v,i}(n)|\forall v\}$ for server $i$.     

        \STATE \textbf{Stage I: Demand-Driven Allocation (All Servers)}
        
        \STATE Initialize the pending vessel set $\mathcal{V}_p = \mathcal{V}$, the assigned vessel set $\mathcal{V}_a = \emptyset$, and the remaining computing resources $F_i^{re}(n) = F_i(n)$.

        \WHILE{$\mathcal{V}_p \neq \emptyset $}
        
            \STATE Allocate computing resource to each vessel $v \in \mathcal{V}_p $ based on $F_{v,i}(n) = \frac{Q_{v,i}(n) / r_{v,i}^c(n)} {\sum_{v \in \mathcal{V}_p} Q_{v,i}(n) / r_{v,i}^c(n)}\cdot F_i^{re}(n)$.

            \IF{$F_{v,i}(n) \leq F_{v,i}^{max}(n), \forall v \in \mathcal{V}_p$}

                \STATE Obtain the computing resource allocation strategy $\boldsymbol{F_{v,i}^I}(n) = \{F_{v,i}(n)|\forall v\}$ of Stage I.
                \STATE \textbf{break}

            \ELSE

                \FOR{$v \in \mathcal{V}_p$ such that $F_{v,i}(n) > F_{v,i}^{max}(n)$}
                    
                    \STATE Set its $F_{v,i}(n) \leftarrow F_{v,i}^{max}(n)$.
                    \STATE Move this vessel $v$ from $\mathcal{V}_p$ to $\mathcal{V}_a$.
    
                \ENDFOR

                \STATE Update $F_i^{re}(n) \leftarrow  F_i(n) - \sum_{v \in \mathcal{V}_a} F_{v,i}(n)$.

            \ENDIF

        \ENDWHILE

        \STATE \textbf{Stage II: Energy-Constrained Scaling (UAV Only)}

        \IF{(31c) is satisfied by $\boldsymbol{F_{v,u}^I}(n)$}
            \STATE Obtain the solution of this UAV $u$ $\boldsymbol{F_{v,u}}(n) = \boldsymbol{F_{v,u}^I}(n)$.
        \ELSE
            \STATE Obtain the scaling coefficient $\beta^*$ via the bisection method such that the scaled solution $\beta^* \cdot \boldsymbol{F_{v,u}^I}(n)$ satisfies (31c) with equality.

            \STATE The final solution $\boldsymbol{F_{v,u}}(n) = \beta^* \cdot \boldsymbol{F_{v,u}^I}(n)$.

        \ENDIF

    \end{algorithmic}
\end{algorithm}

As the final stage of task execution in SAGSIN, computing resource management governs the terminal processing rate for each vessel's task data. Each server should adaptively allocate computing resources based on the backlog and computational demands of different vessels as well as its real-time computing capability, which is crucial for mitigating system-wide task congestion and accelerating task completion.

To this end, we devise a two-stage computing resource allocation scheme for heterogeneous servers while satisfying the relevant constraints, i.e., (16e)-(16i). Denoting the computing resource that server $i$ allocates to vessel $v$ at time slot $n$ as $F_{v,i}(n)$, measured in CPU cycles, the above constraints can be rewritten as
\begin{align*}
& \sum_{v \in \mathcal{V}} F_{v,i}(n) \leq F_i(n) , \forall i, \tag{31a}\\
& 0 \leq F_{v,i}(n) \leq F_{v,i}^{max}(n), \forall v,i, \tag{31b}\\
& \sum_{v \in \mathcal{V}} \frac{\kappa_u F_{v,u}(n)^3}{\tau^2} -\frac{P_u(n+1)}{R_{u,h}(n+1)} \cdot \sum_{v \in \mathcal{V}} \frac{F_{v,u}(n)} {C_v} \leq A_u^5(n), \forall u, \tag{31c}
\end{align*}
where $A_u^5(n) = E_u(n) - E_u^{tx}(n) - E_u^{flight}(n) - E_u^{cs}(n+1) - \frac{P_u(n+1)} {R_{u,h}(n+1)} \cdot \sum_{v \in \mathcal{V}} (Q_{v,u}(n) + a_{v,u}(n) \cdot D_{v,u}^{tx}(n) - a_{v,u,h}(n) \cdot D_{v,u,h}^{tx}(n) )$ and we denote the maximum required resources of vessel $v$ at server $i$ as $F_{v,i}^{max}(n) = C_v \cdot ( Q_{v,i}(n) + D_{v,i}^{tx}(n) - \boldsymbol{1}_{\{ i \in \{u,h\} \}} \cdot D_{v,i,i'}^{tx}(n) ), \forall i $. The proposed computing resource allocation scheme for each server is detailed in \textbf{Algorithm~\ref{alg:computing}}. Only the key parameters of this subsection are listed in the input, while other required parameters are omitted for concise presentation. In Stage I, the computing resources of each server are allocated in proportion to the completion time of each vessel's data backlog on this server while satisfying (31a) and (31b), thereby prioritizing tasks with severe backlog. For each UAV server, Stage II is further executed. If the solution obtained in Stage I violates (31c), it will be scaled in Stage II to ensure the constraint is satisfied. After obtaining the final solution, $D_{v,i}^{comp}(n) = F_{v,i}^*(n) / C_v , \forall v,i$.


\subsection{Complexity Analysis}

The proposed overall approach is summarized in \textbf{Algorithm~\ref{alg:overall}}. At the beginning, we initialize the system parameters and update data backlog queues of all vessels and servers. Then \textbf{Algorithm~\ref{alg:offloading}} is first executed to obtain the multi-layer task offloading decisions. Subsequently, according to the dependencies among the subproblems, we iteratively optimize the UAV-BS bandwidth allocation, UAV trajectory planning, and computing resource allocation for heterogeneous servers in sequence, until convergence or the maximum number of iterations is reached. The superscript $j$ denotes the solution at the start of the $j$-th iteration.

\begin{algorithm}[!t]
    \caption{Proposed Overall Method of Dynamic Task and Resource Scheduling for SAGSIN.}
    \label{alg:overall}
    \renewcommand{\algorithmicrequire}{\textbf{Input:}}
    \renewcommand{\algorithmicensure}{\textbf{Output:}}
    
    \begin{algorithmic}[1]
        \REQUIRE The UAV positions $\boldsymbol{W_u}(n-1)$ in previous time slot and all the remaining parameters in current time slot.   
        \ENSURE Task offloading decisions $\boldsymbol{a}(n)$, UAV-BS bandwidth allocation $\boldsymbol{B_b}(n)$ and $\boldsymbol{B_u}(n)$, computing resource allocation $\boldsymbol{D^{comp}}(n)$, and UAV trajectories
        $\boldsymbol{W_u}(n)$.
        
        \STATE Update data backlog for all devices according to (7) or set $Q_{v,0}(0) = D_v$ and $Q_{v,i}(0) = 0$ at the initial time slot. Initialize the iteration number $j=0$.
        \STATE Execute \textbf{Algorithm~\ref{alg:offloading}} to obtain $\boldsymbol{a}(n)$.

        \REPEAT
            \STATE Solve $\mathcal{P}_1$ based on (23) to obtain $\boldsymbol{B_b}(n)^{j+1}$ and $\boldsymbol{B_u}(n)^{j+1}$ with given $\boldsymbol{W_u}(n)^j$ and $\boldsymbol{D^{comp}}(n)^j$.
            
            \STATE Solve $\mathcal{P}_2'$ to obtain $\boldsymbol{W_u}(n)^{j+1}$ with given $\boldsymbol{B_u}(n)^{j+1}$ and $\boldsymbol{D^{comp}}(n)^j$.
            
            \STATE Execute \textbf{Algorithm~\ref{alg:computing}} to obtain $\boldsymbol{D^{comp}}(n)^{j+1}$ for all servers with given $\boldsymbol{B_b}(n)^{j+1}$, $\boldsymbol{B_u}(n)^{j+1}$, and $\boldsymbol{W_u}(n)^{j+1}$.
            
            \STATE Update the value of objective function.
            
            \STATE Update $j=j+1$.

        \UNTIL The objective value converges or $j>j_{max}$.

    \end{algorithmic}
\end{algorithm}

In \textbf{Algorithm~\ref{alg:offloading}}, calculating the rates and distances for all links requires two traversals of them, involving a complexity of $\mathcal{O}\left( 2(2VU+4V+U+1) \right) \approx \mathcal{O} \left(VU\right)$. To calculate the shortest distance, each UAV and the HAP only need to compare 3 and 2 paths for each vessel, respectively, whereas the shortest distances for the BS and satellite are unique. Each vessel needs to compare $3U+1$ paths. Thus, the complexity of obtaining PI is $\mathcal{O}\left( V(3U+1) + 3VU + 2V \right) \approx \mathcal{O} \left(VU\right)$. Calculating PD for all V2U/B, U2H, and H2S links involves complexity of $\mathcal{O}\left( V(U+1) + VU+ V \right) \approx \mathcal{O} \left(VU\right)$. Since the multi-layer offloading decisions are made by comparing the PD of each vessel’s data across all communication links, this process has the same complexity as PD calculation. Therefore, the overall complexity of \textbf{Algorithm~\ref{alg:offloading}} is only $\mathcal{O}\left( 4VU \right) \approx \mathcal{O} \left(VU\right)$. $\mathcal{P}_1$ is solved via closed-form solution and an optional bisection search with tolerance $\epsilon$, yielding a worst-case complexity of $\mathcal{O} \left(V + U\log_2(\frac{1}{\epsilon}) \right)$. Solving $\mathcal{P}_2'$ by SCA method involves a complexity of $\mathcal{O}((VU)^{3.5})$. The worst-case complexities of Stage I and Stage II in \textbf{Algorithm~\ref{alg:computing}} are $\mathcal{O}(V^2)$ and $\mathcal{O} \left( V\log_2(\frac{1}{\epsilon})\right)$, respectively, yielding an overall complexity of $\mathcal{O} \left( V^2U + VU\log_2(\frac{1}{\epsilon}) + 
3V^2 \right) \approx \mathcal{O} \left( V^2U \right)$ for computing resource allocation of all servers. Therefore, the complexity of the overall proposed method summarized in \textbf{Algorithm~\ref{alg:overall}} is given by $\mathcal{O} \left( VU + J ( V + U\log_2(\frac{1}{\epsilon}) + (VU)^{3.5} + V^2U ) \right) \approx \mathcal{O} \left( J(VU)^{3.5} \right)$, where $J$ denotes the number of iterations. Despite the inherently exponential complexity of multi-layer task offloading, the proposed method further incorporates multifaceted resource optimization while maintaining only polynomial complexity. This demonstrates its remarkable computational efficiency for timely resource scheduling under massive connectivity in SAGSIN.


\section{Simulation Results}

In this section, we conduct extensive simulation experiments to validate the effectiveness of the proposed method through comparisons with benchmarks. Then we further evaluate its performance in terms of load balancing, UAV energy management, and handover backlog control.

\subsection{Scenario}

We consider a 2 km$\times$2 km ocean area. The lower-left corner of the considered area is set as the origin of the horizontal coordinate system. Six UAVs are initially deployed at (0.5 km, 0.5 km), (0.5 km, 1 km), (0.5 km, 1.5 km), (1.5 km, 0.5 km), (1.5 km, 1 km), and (1.5 km, 1.5 km), with the maximum speed of 15 m/s and minimum safety distance of 5 m \cite{sqi2025joint}. The vessel locations follow the Poisson distribution, and each travels along a randomly generated trajectory with the velocity ranging from 5 to 15 m/s \cite{wwu2025multi}. The UAV charging station and coastal BS are located at (0, 0.9 km) and (0, 1 km), respectively, while the HAP remains quasi-stationary at the center of the region \cite{wli2026efficient}. Similar to \cite{zwang2025double, wwu2025multi}, the settings of remaining key parameters are summarized in Table~\ref{tab:simulation}. The values of bandwidth and computing resources listed in Table~\ref{tab:simulation} serve as the default settings, which are varied around these baseline values in subsequent experiments when required.


\subsection{Benchmarks}

To validate the effectiveness of the proposed DASH method, we compare it with the following benchmark schemes:

\begin{itemize}
    \item \textbf{DASH w/o HO}: The proposed DASH method without anticipatory satellite handover strategy.
    
    \item \textbf{HACO} \cite{wwu2025multi}: A multi-\underline{H}AP-\underline{a}ssisted \underline{c}omputation \underline{o}ffloading algorithm for SAGSIN aiming at minimizing the overall task execution delay.
    
    \item \textbf{FLEC} \cite{xwang2024amtos}: A \underline{f}our-\underline{l}ayer \underline{e}dge \underline{c}omputing scheme that aims to minimize system task costs within a similar hierarchical architecture to this study.

    \item \textbf{SHAR} \cite{rwibrahim2025latency}: A \underline{s}atellite-\underline{HA}P collaborative computing algorithm with \underline{r}eactive handover scheme, where the full data of unfinished tasks on the current satellite at handover are reoffloaded to the incoming satellite.
    
\end{itemize}



\begin{table}[!t]
\caption{Simulation Parameters}
\label{tab:simulation}
\centering
\begin{tabular}{|m{1.6cm}<{\centering}|m{1.9cm}<{\centering}|m{1.8cm}<{\centering}|m{1.7cm}<{\centering}|}
\hline
\textbf{Parameter} & \textbf{Value} & \textbf{Parameter} & \textbf{Value}\\
\hline
$V$ & 5--30 & $N_0$ & -174 dBm/Hz\\
\hline
$D_v$ & 2--10 Mb & $H_u, H_b, H_h$ & [0.1, 0.03, 20] km\\
\hline
$C_v$ & 100--2000 & $d_{h,s}$ & 784 km\\
\hline
$F_u, F_b, F_h,F_s$ & [1, 3, 3, 10] ×$10^8$ cycles/slot & $B_u, B_b, B_h, B_s$ & [10, 20, 20, 50] MHz\\
\hline
$P_v, P_u, P_h$ & [1, 1.5, 2.5] W & $L_0$ & -30 dB\\
\hline
$L_g, L_s$ & [23, 0] dB & $\mu, \hat{\mu}$ & [2, 46.4]\\
\hline
$G_u, G_h, G_s$ & [25, 30, 35] dBi & $T_p, \tau$ & [0.01, 0.1] s\\
\hline

\end{tabular}
\end{table}

\begin{figure}[!t]
\centerline{\includegraphics[width=2.8 in]{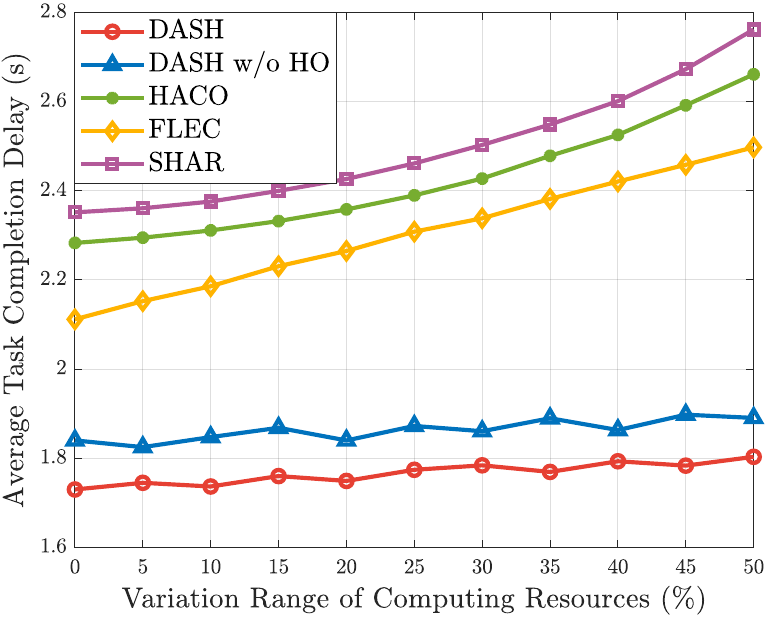}}
\caption{The comparison of average task completion delay among different methods with varying
fluctuation range of available computing resources.}
\label{exp_comp_var}
\end{figure}

\subsubsection{Average task delay under varying computing resources}

Fig.~\ref{exp_comp_var} illustrates the average task completion delay among different methods with the percentage variation range of available computing resources of servers across time slots. DASH adaptively adjusts the fine-grained task and resource scheduling strategy based on the real-time computing resources of all servers and the incoming satellite, consistently achieving the lowest average delay and the slowest growth trend. When the incoming satellite has limited computing capacity, DASH w/o HO cannot proactively constrain the satellite offloading volume, leading to post-handover congestion and increased satellite computation delay. Conversely, when the incoming satellite has more computing resources, it also fails to adaptively increase satellite offloading volume to exploit. The one-shot task-resource scheduling of HACO, FLEC and SHAR leads to inconsistency between the available computing resources at the decision-making time and the time when tasks reach destination servers. Upon task arrival, reduced computing resources increase computation delay, whereas increased resources cannot be exploited. Consequently, their average delays grow monotonically with intensifying computing resource fluctuations. For SHAR, the entire tasks that are unfinished on the current satellite at handover need to be recomputed on the incoming satellite, further increasing their computation delay and sensitivity to computing resource variations. DASH adaptively routes task data based on instantaneous server states, which effectively absorbs the impact of resource fluctuations, reducing the average delay by around 30\%, 28\%, and 24\% compared to SHAR, HACO, and FLEC, respectively.

\begin{figure}[!t]
\centerline{\includegraphics[width=2.8 in]{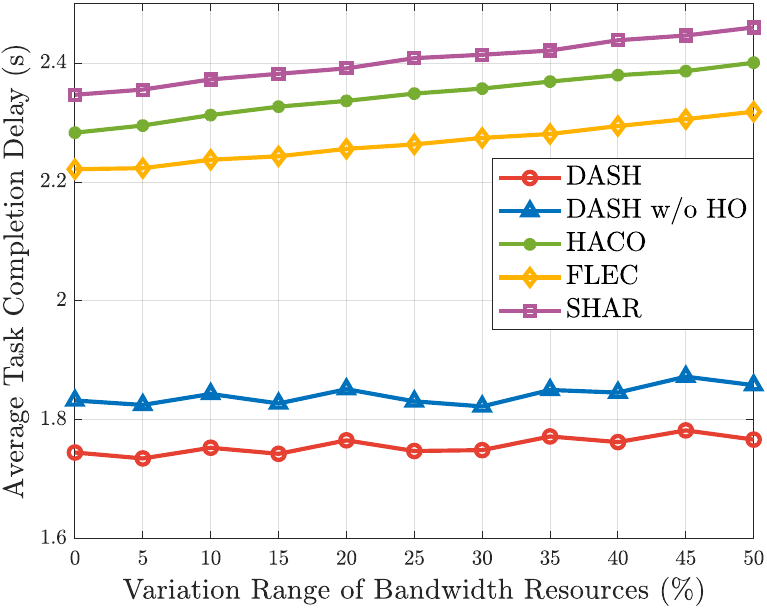}}
\caption{The comparison of average task completion delay among different methods with varying
fluctuation range of available bandwidth resources.}
\label{exp_bw_var}
\end{figure}

\subsubsection{Average task delay under varying bandwidth resources}

Fig.~\ref{exp_bw_var} depicts the average task completion delay among different methods with the percentage variation range of available bandwidth for wireless links in the system across time slots. For the proposed DASH method and its ablation variant DASH w/o HO, task offloading and bandwidth allocation are dynamically adjusted based on real-time link states, including available bandwidth, large and small-scale fading, and channel capacities, enabling strong adaptability to communication dynamics. DASH reduces the average task delay respectively by around 27\%, 25\%, and 23\% compared to SHAR, HACO, and FLEC, since their one-shot scheduling strategies fail to adaptively reroute task when channel conditions deteriorate or reallocate resources when bandwidth becomes abundant. The reactive satellite handover mechanism of SHAR requires unfinished satellite tasks to be retransmitted to incoming satellite, which amplifies the communication latency under limited bandwidth resources. Compared with more dominant computing resources, which govern the terminal processing rates, the ‌intensification‌ of bandwidth fluctuations has a relatively smaller impact on their average delay increment.

\begin{figure}[!t]
\centerline{\includegraphics[width=2.8 in]{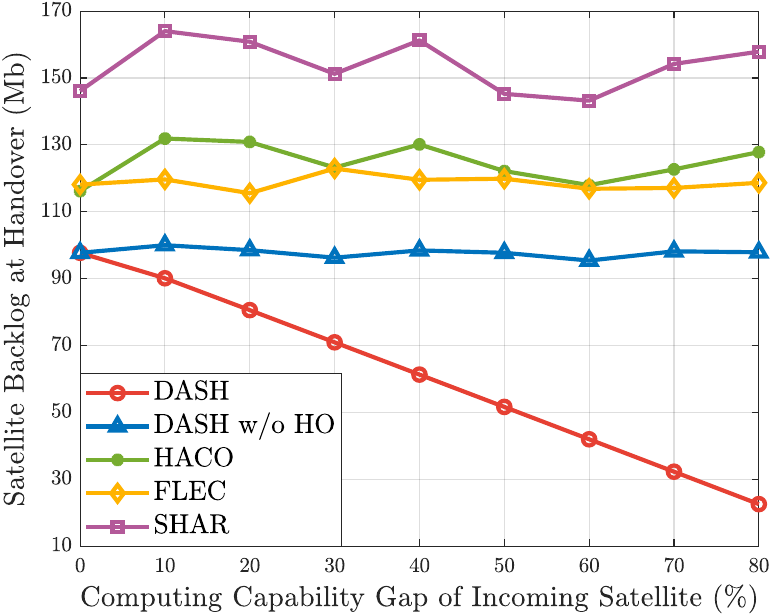}}
\caption{The comparison of satellite backlog data at handover among different methods with varying
computing resource gap of incoming satellite.}
\label{exp_sat_backlog}
\end{figure}

\subsubsection{Handover backlog with different computing resource gap of incoming satellite}

Fig.~\ref{exp_sat_backlog} shows the amount of satellite backlog at handover (i.e., task data volume received by the incoming satellite after handover) versus the deficit percentage in the average available computing resources of the incoming satellite relative to the current one. Benefiting from the anticipatory handover strategy, DASH adaptively restricts the satellite offloading volume as the incoming satellite’s computing capacity continuously decreases, thereby markedly reducing satellite backlog at handover and alleviating post-handover congestion. With fine-grained and balanced task-resource scheduling, DASH w/o HO achieves lower satellite backlog at handover than SHAR, HACO, and FLEC. Due to the reactive handover mechanism of SHAR, the entire data of unfinished tasks at handover on the current satellite are reoffloaded to the incoming satellite, including the already processed portions. This increases the post-handover satellite backlog accumulation. However, since these four methods determine offloading strategies solely based on the current satellite state, they lack adaptability to changes in the computing capability of incoming satellite. Consequently, their handover backlogs remain insensitive to the increasing gaps of the incoming satellite’s computing resources.

\begin{figure}[!t]
\centerline{\includegraphics[width=2.8 in]{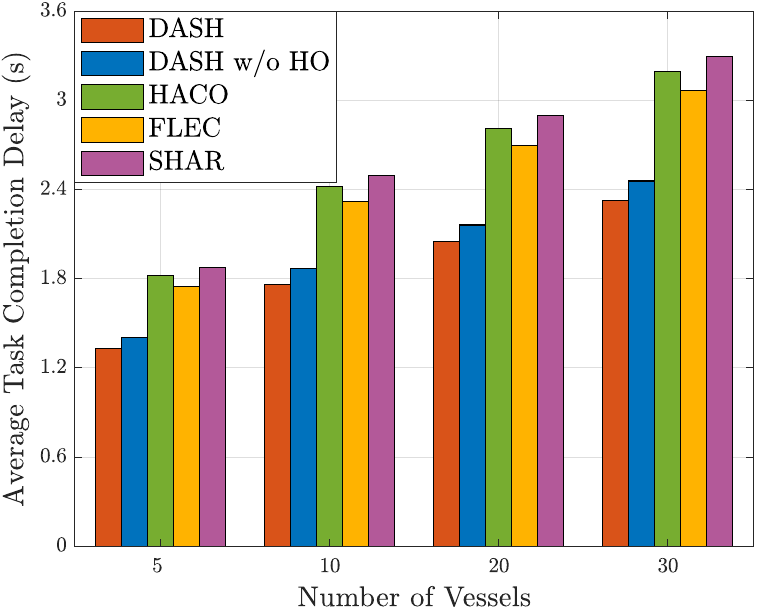}}
\caption{The comparison of average task completion delay among different methods with varying number of vessels.}
\label{exp_vessel_num}
\end{figure}

\subsubsection{Average task delay with varying number of vessels}

Fig.~\ref{exp_vessel_num} illustrates the average task completion delay among different methods under varying number of vessels. As the number of vessels increases, all methods experience higher delay due to the growing task demands and intensified competition for communication and computing resources. DASH consistently achieves the lowest delay and exhibits the slowest growth trend, validating its superior scalability under different network scales. With fine-grained task routing based on real-time server states, DASH effectively balances workloads among heterogeneous servers and mitigates congestion caused by increasing access demands. HACO, FLEC, and SHAR establish task offloading strategies according to initial network states, which remain fixed during subsequent task execution and thus limit the efficient utilization of system resources under increasing task demands. Additionally, the binary offloading mechanisms of HACO and SHAR may overload certain servers under dense vessel scenarios, further increasing task execution delay.


\subsection{Performance Analysis of the Proposed Method}

\begin{figure}[!t]
\centerline{\includegraphics[width=2.8 in]{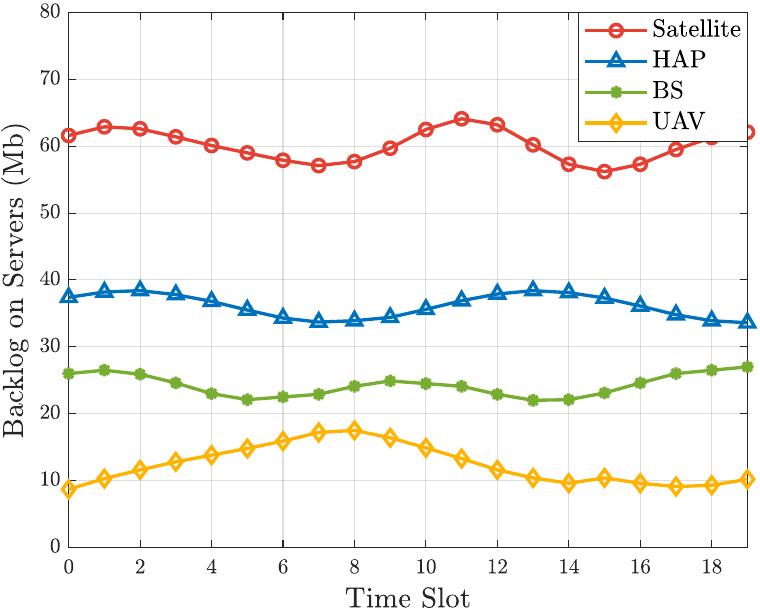}}
\caption{The task backlog on different types of servers across time slots.}
\label{exp_server_backlog}
\end{figure}

\subsubsection{Task backlog among heterogeneous servers}

Fig.~\ref{exp_server_backlog} illustrates the task backlog evolution on different types of servers during selected time slots of system operation. The backlog levels are not identical across servers, which reflects the heterogeneous roles and service capabilities of different layers. As a broad-coverage, high-capacity server, the satellite undertakes a relatively large workload. The HAP and BS, which mainly serve offshore and nearshore areas, respectively, act as medium-scale servers and carry moderate task loads, while UAVs mainly function as flexible low-altitude relay nodes and thus bear lighter burdens. Furthermore, the backlog queues of servers in all layers remain dynamically stable, and their backlog peaks and valleys are generally staggered. This effectively validates the cross-layer load-balancing capability of DASH enabled by its joint adaptation mechanism of task and communication-computing resource scheduling. For task scheduling, DASH dynamically updates the PI of each server in each time slot. Severe backlog at a server will increase its PI, which reduces its offloading attractiveness in subsequent time slots and thereby suppresses the inflow of new task data. In terms of communication, DASH flexibly coordinates bandwidth resources based on adaptive task scheduling, which promotes efficient task transmission towards servers with lighter workloads and abundant computing resources. Regarding task computation, DASH allocates computing resources according to the computational demands of different vessels and their task data buffered on each server, further accelerating backlog processing in a demand-driven manner. 


\begin{figure}[!t]
\centerline{\includegraphics[width=3.0 in]{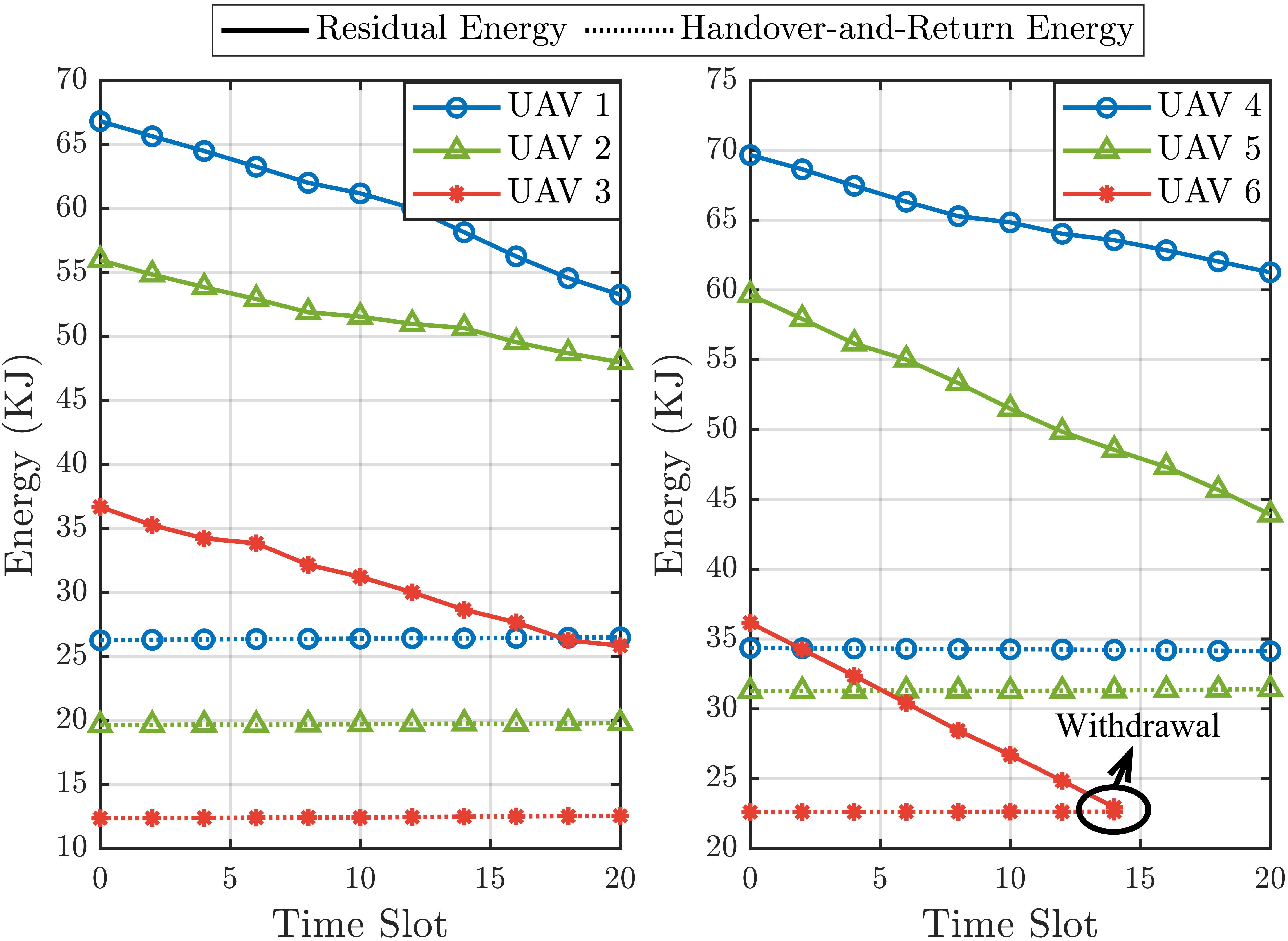}}
\caption{The residual energy and handover-and-return energy of UAV servers.}
\label{exp_uav_energy}
\end{figure}

\subsubsection{UAV energy management}

Fig.~\ref{exp_uav_energy} depicts the trends of the residual energy and the energy required for task backlog handover and safe return (termed handover-and-return energy) of UAV servers during selected time slots of system operation. For each UAV, its residual energy gradually decreases due to data transmission, task computation, and flight operations, while the required handover-and-return energy fluctuates at a relatively low level due to variations in UAV positions and buffered task backlog. It can be observed that the residual energy of each UAV consistently remains above its handover-and-return energy. When the residual energy of a certain UAV approaches the safety threshold (e.g., the UAV 6), it withdraws from service and returns to the charging station, after which a newly dispatched UAV will take over its role. This validates the effectiveness of the proposed energy-aware UAV scheduling design, which ensures UAV operational safety and service continuity.

\begin{figure}[!t]
\centerline{\includegraphics[width=2.8 in]{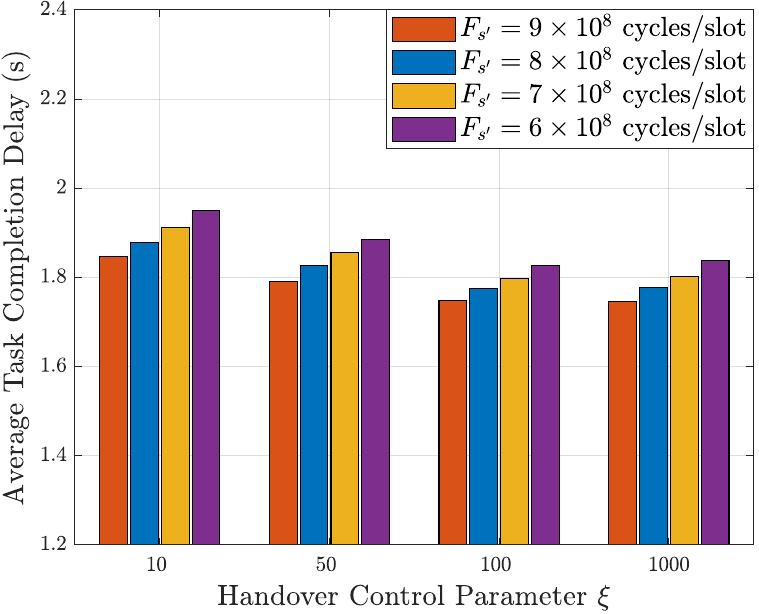}}
\caption{Impact of the handover control parameter $\xi$ on the average task delay under varying computing resource gap of incoming satellite.}
\label{exp_xi}
\end{figure}

\subsubsection{Impact of pre-handover period}

Fig.~\ref{exp_xi} shows the influence of the handover control parameter $\xi$ on the average task delay under different computing capability gaps between the current and incoming satellites. $\xi$ determines the length of pre-handover period during which DASH accounts for the incoming satellite state. A larger $\xi$ leads to a longer pre-handover period, allowing DASH to regulate satellite offloading traffic based on both satellite states earlier. Given the same $\xi$, a larger capability gap results in higher task delay due to the reduced satellite computing resources after handover. For a certain gap, when $\xi$ increases from 10 to 100, the average task delay decreases because a longer preparation period enables DASH to incorporate the incoming satellite state earlier and control the data volume offloaded to satellite. This effectively alleviates post-handover congestion and thereby reduces the average delay. However, further increasing $\xi$ from 100 to 1000 brings little additional benefit and may even slightly increase the delay. This is because an excessively long pre-handover period causes the satellite offloading strategy to be affected by the incoming satellite state too early. As a result, DASH may underutilize the current satellite computing power and restrict satellite offloading conservatively. This indicates that the control parameter should be properly configured according to practical system requirements, while DASH consistently provides effective adaptation under different settings.

\section{Conclusions}

This paper proposed a dynamic task and resource scheduling method for SAGSIN to minimize task execution delay of vessel users. To address the complexity and high dynamics of multi‑layer task offloading, we developed a layer‑wise offloading scheme that adapts to real‑time system states, along with an anticipatory satellite handover policy to mitigate post‑handover congestion. Furthermore, we jointly optimized UAV‑BS bandwidth allocation, UAV trajectories, and computing resources to accelerate task completion and enhance resource utilization, while ensuring adequate residual energy for backlog handover and safe return of UAVs. Experimental results show that the proposed method significantly reduces task delay under fluctuating system resources and adaptively controls satellite backlog at handover compared with benchmarks. This study provides important insights for the development of efficient and sustainable SAGSIN systems.


 




\vfill


\begin{thebibliography}{10}
\bibliographystyle{IEEEtran}

\bibitem{majamshed2025non}
M. A. Jamshed et al., ``Non-terrestrial networks for 6G: Integrated, intelligent, and ubiquitous connectivity," \textit{IEEE Commun. Standards Mag.}, vol. 9, no. 3, pp. 86-93, Sept. 2025.


\bibitem{jyou2025joint}
J. You, Z. Jia, C. Dong, Q. Wu, and Z. Han, ``Joint computation offloading and resource allocation for uncertain maritime MEC via cooperation of AAVs and vessels," \textit{IEEE Trans. Veh. Technol.}, vol. 74, no. 11, pp. 18081-18095, Nov. 2025.


\bibitem{zhuang2026joint}
Z. Huang, Z. Yu, L. Wang, H. Zhou, and B. Guo, ``Joint optimization of caching, migration, and offloading in satellite-assisted marine networks," \textit{IEEE Trans. Netw.}, vol. 34, pp. 5082-5097, 2026. 


\bibitem{zwang2025two}
Z. Wang, B. Lin, Q. Ye, and H. Peng, ``Two-tier task offloading for satellite-assisted marine networks: A hybrid stackelberg–bargaining game approach," \textit{IEEE Internet Things J.}, vol. 12, no. 9, pp. 13047-13060, 1 May1, 2025. 


\bibitem{zhuang2026two}
Z. Huang, Z. Yu, L. Wang, Y. Zhao, H. Zhou, and B. Guo, ``Two time-scale DRL for service caching and task offloading in cross-domain marine networks," \textit{IEEE Trans. Mobile Comput.}, vol. 25, no. 1, pp. 785-800, Jan. 2026.


\bibitem{yhe2025delay}
Y. He, F. Huang, D. Wang, L. Yang, and R. Zhang, ``Delay minimization for NOMA-MEC offloading in ABS-aided maritime communication networks," \textit{IEEE Trans. Veh. Technol.}, vol. 74, no. 6, pp. 9577-9590, June 2025. 


\bibitem{yye2026multitier}
Y. Ye, S. Gao, X. Zheng, and L. Yang, "Multi-tier UAV edge computing toward long-term energy stability for low altitude networks," \textit{IEEE Open J. Commun. Soc.}, vol. 7, pp. 6090-6104, 2026.


\bibitem{ybai2025dynamic}
Y. Bai and Y. Zhang, ``Dynamic offloading based on Lyapunov optimization for UAV-assisted maritime IoT-MEC networks," \textit{IEEE Trans. Veh. Technol.}, vol. 74, no. 11, pp. 17894-17906, Nov. 2025.


\bibitem{jning2025marho}
J. Ning, A. Li, G. C. F. Lee, S. Sun, and T. Yang, ``MARHO: Hybrid task offloading in maritime MEC via multi-agent reinforcement learning," \textit{IEEE Open J. Commun. Soc.}, vol. 6, pp. 10322-10337, 2025.


\bibitem{wxu2025madrl}
W. Xu, W. Luo, Y. Sun, Z. Gao, B. Wu, and L. Lai, ``MADRL-based edge computing: Joint energy-latency optimization for marine internet of things," \textit{IEEE Internet Things J.}, vol. 12, no. 15, pp. 30228-30241, 1 Aug.1, 2025. 

\bibitem{mdai2023uav}
M. Dai, Y. Wu, L. Qian, Z. Su, B. Lin, and N. Chen, ``UAV-assisted multi-access computation offloading via hybrid NOMA and FDMA in marine networks," \textit{IEEE Trans. Netw. Sci. Eng.}, vol. 10, no. 1, pp. 113-127, 1 Jan.-Feb. 2023.


\bibitem{sqi2024minimizing}
S. Qi, B. Lin, Y. Deng, X. Chen, and Y. Fang, ``Minimizing maximum latency of task offloading for multi-UAV-assisted maritime search and rescue," \textit{IEEE Trans. Veh. Technol.}, vol. 73, no. 9, pp. 13625-13638, Sept. 2024. 


\bibitem{sgao2026integrated}
S. Gao et al., ``Integrated sensing, communication, and computation for low-altitude networks towards seamless connectivity and connected intelligence," \textit{IEEE Internet Things Mag.}, vol. 9, no. 3, pp. 63-71, May 2026.


\bibitem{xwang2025bridging}
X. Wang et al., ``Bridging terrestrial and non-terrestrial networks: A novel architecture for space-air-ground-sea integration system," \textit{IEEE Wireless Commun.}, vol. 32, no. 3, pp. 20-27, June 2025.


\bibitem{mdai2023latency}
M. Dai et al., ``Latency minimization oriented hybrid offshore and aerial-based multi-access computation offloading for marine communication networks," \textit{IEEE Trans. Commun.}, vol. 71, no. 11, pp. 6482-6498, Nov. 2023.


\bibitem{yzhang2025joint}
Y. Zhang, Z. Na, S. Li, B. Lin, Y. Lin, and A. Nallanathan, ``Joint service caching and task offloading for multi-UAV-assisted offshore edge computing networks," \textit{IEEE Trans. Veh. Technol.}, vol. 74, no. 12, pp. 19667-19680, Dec. 2025.


\bibitem{mli2026online}
M. Li, L. P. Qian, F. Fang, and X. Wang, ``Online hierarchical computation offloading for marine IoT networks: A delay minimization approach," \textit{IEEE Trans. Wireless Commun.}, vol. 25, pp. 3422-3436, 2026.


\bibitem{zwang2025double}
Z. Wang, B. Lin, and Q. Ye, ``Double-edge-assisted computation offloading and resource allocation for space-air-marine integrated networks," \textit{IEEE Trans. Veh. Technol.}, vol. 74, no. 9, pp. 14501-14514, Sept. 2025.


\bibitem{sjung2023marine}
S. Jung, S. Jeong, J. Kang, and J. Kang, ``Marine IoT systems with space–air–sea integrated networks: Hybrid LEO and UAV edge computing," \textit{IEEE Internet Things J.}, vol. 10, no. 23, pp. 20498-20510, 1 Dec.1, 2023.


\bibitem{sqi2025joint}
S. Qi, B. Lin, Y. Deng, H. Pan, and X. Hu, ``Joint computation offloading and resource management for cooperative satellite–aerial–marine internet of things networks," \textit{IEEE Internet Things J.}, vol. 12, no. 24, pp. 53164-53176, 15 Dec.15, 2025.


\bibitem{mdai2025energy}
M. Dai, S. Chang, Y. Wang, and Z. Su, ``Energy-efficient multi-access edge computing for heterogeneous satellite-maritime networks: A hybrid harvesting-and-offloading design," \textit{IEEE Trans. Mobile Comput.}, vol. 24, no. 11, pp. 12001-12018, Nov. 2025.


\bibitem{hzhang2025energy}
H. Zhang, S. Xi, B. Shang, P. Zhang, S. Wu, and C. Jiang, ``Energy oriented three-tier computation offloading scheme in maritime edge computing network," \textit{IEEE Trans. Veh. Technol.}, vol. 74, no. 5, pp. 8126-8140, May 2025.


\bibitem{tyang2022multi}
T. Yang et al., ``Multi-armed bandits learning for task offloading in maritime edge intelligence networks," \textit{IEEE Trans. Veh. Technol.}, vol. 71, no. 4, pp. 4212-4224, April 2022.


\bibitem{wwu2025multi}
W. Wu, W. Feng, Y. Fang, Z. Lin, and X. Lu, ``Multi-HAP-assisted computation offloading in space–air–ground–sea integrated network," \textit{IEEE Internet Things J.}, vol. 12, no. 12, pp. 21806-21818, 15 June15, 2025.


\bibitem{wli2026efficient}
W. Li, S. Li, J. Hao, Q. Wu, and R. Wang, "Efficient task offloading and resource allocation in HAPS-assisted LEO satellite networks: A MAPPO with exact potential game approach," \textit{IEEE Internet Things J.}, vol. 13, no. 6, pp. 10179-10195, 15 March15, 2026.


\bibitem{zlin2024maritime}
Z. Lin, J. Yang, Y. Chen, C. Xu, and X. Zhang, ``Maritime distributed computation offloading in space-air-ground-sea integrated networks," \textit{IEEE Commun. Lett.}, vol. 28, no. 7, pp. 1614-1618, July 2024.


\bibitem{dwang2023double}
D. Wang, T. He, Y. Lou, L. Pang, Y. He, and H. -H. Chen, ``Double-edge computation offloading for secure integrated space–air–aqua networks," \textit{IEEE Internet Things J.}, vol. 10, no. 17, pp. 15581-15593, 1 Sept.1, 2023.


\bibitem{sshassan2021blue}
S. S. Hassan, Y. K. Tun, W. Saad, Z. Han, and C. S. Hong, ``Blue data computation maximization in 6G space-air-sea non-terrestrial networks," in \textit{Proc. IEEE Glob. Commun. Conf. (GLOBECOM)}, Madrid, Spain, 2021, pp. 1-6.


\bibitem{gwang2024free}
G. Wang, F. Yang, J. Song, and Z. Han, ``Free space optical communication for inter-satellite link: Architecture, potentials and trends," \textit{IEEE Commun. Mag.}, vol. 62, no. 3, pp. 110-116, March 2024.


\bibitem{resswein2026constellation}
R. Esswein, Q. Bayer, S. Mergendahl, J. Ruffley, M. Abdelhakim, and R. K. Cunningham, ``Constellation parameters for minimizing propagation delay over LEO inter-satellite links," \textit{IEEE Trans. Netw.}, doi: 10.1109/TON.2026.3661441.


\bibitem{xdeng2023distance}
X. Deng, L. Chang, S. Zeng, L. Cai, and J. Pan, ``Distance-based back-pressure routing for load-balancing LEO satellite networks," \textit{IEEE Trans. Veh. Technol.}, vol. 72, no. 1, pp. 1240-1253, Jan. 2023.


\bibitem{she2025back}
S. He, Y. He, Y. Zhang, Q. Shi, and T. Z. Qiu, ``Back-pressure-based traffic signal and discretized trajectory joint control for low CAV penetration rate environment," \textit{IEEE Trans. Intell. Veh.}, vol. 10, no. 4, pp. 2679-2697, April 2025.


\bibitem{jzhang2026energy}
J. Zhang et al., ``Energy-efficient UAV deployment and computation offloading in space-air-ground integrated networks," \textit{IEEE Trans. Veh. Technol.}, vol. 75, no. 2, pp. 3081-3098, Feb. 2026.


\bibitem{xwang2024amtos}
X. Wang, S. Wang, X. Gao, Z. Qian, and Z. Han, ``AMTOS: An ADMM-based multilayer computation offloading and resource allocation optimization scheme in IoV-MEC system," \textit{IEEE Internet Things J.}, vol. 11, no. 19, pp. 30953-30964, 1 Oct.1, 2024.


\bibitem{rwibrahim2025latency}
R. W. Ibrahim, T. K. Rodrigues, N. Kato, M. Ariyoshi, and Y. Hasegawa, ``Latency-aware task offloading in multi-tier SAGIN with FSO-enabled mobile edge computing," \textit{IEEE Trans. Veh. Technol.}, vol. 75, no. 6, pp. 10368-10380, June 2026.








%










\end{thebibliography}
\end{document}